\documentclass[twocolumn,showpacs,amsmath,amssymb,superscriptaddress,pre,floatfix]{revtex4}

\usepackage{amsmath,amssymb,latexsym,graphicx,epsfig}
\usepackage{pstricks,psfrag}
\newpsobject{showgrid}{psgrid}{subgriddiv=1}

\newcommand{\td}{\mathrm{d}}

\newcommand{\be}{\begin{equation}}
\newcommand{\ee}{\end{equation}}
\newcommand{\bea}{\begin{eqnarray}}
\newcommand{\eea}{\end{eqnarray}}
\numberwithin{equation}{section}

\begin{document}

\title{Renormalization flow in  extreme value statistics}

\author{Eric Bertin}\email{eric.bertin@ens-lyon.fr}
\affiliation{Universit\'e de Lyon, Laboratoire de Physique, ENS Lyon,
CNRS,
46 All\'ee d'Italie, F-69007 Lyon, France}

\author{G\'eza Gy\"{o}rgyi}
\email{gyorgyi@glu.elte.hu}
\affiliation{Department of Theoretical Physics, University of Geneva,
Geneva, Switzerland \\ and Department of Materials Physics,
E\"{o}tv\"{o}s University,  Budapest, Hungary}

\date{\today}

\begin{abstract}

The renormalization group transformation for extreme value statistics of
independent, identically distributed variables, recently introduced to describe
finite size  effects, is presented here in terms of a partial differential
equation (PDE).  This yields a flow in function  space and gives rise  to the
known family of Fisher-Tippett limit distributions as fixed points, together
with the universal eigenfunctions around them. The PDE turns out to handle
correctly distributions  even  having discontinuities.  Remarkably, the PDE
admits exact solutions in terms of eigenfunctions even farther from the fixed
points.  In particular, such are unstable  manifolds emanating from and
returning to the Gumbel fixed point, when the running eigenvalue and the
perturbation strength parameter obey a pair of coupled ordinary differential
equations.  Exact renormalization trajectories corresponding to linear
combinations of eigenfunctions can also be given, and it is shown that such are
all solutions of the PDE.  Explicit formulas for some invariant manifolds in the
Fr\'echet and Weibull cases are also presented. Finally, the similarity between
renormalization flows for  extreme value statistics and the central limit
problem is stressed, whence follows the equivalence of the formulas for Weibull
distributions and the moment generating function of symmetric L\'evy stable
distributions.
\end{abstract}


\maketitle
\tableofcontents
%%%%%%%%%%%%%%%%%%%%%%%%%%%%%%%%%%%%%%%%%%%%%%%%%%%%%%%%%%%%%%%%%%%%%%
%%%%%%
\section{Introduction}
\label{S:intro}
%%%%%%%%%%%%%%%%%%%%%%%%%%%%%%%%%%%%%%%%%%%%%%%%%%%%%%%%%%%%%%%%%%%%%%
%%%%

Extreme value statistics (EVS) is a prominent statistical problem, which
attracts the attention of an increasing number of physicists as
well as scientists from other disciplines.  In its traditional proposition
EVS concerns the statistics of the extremal, i.e., largest or smallest value in
a batch of random quantities, or, in general the statistics of an ordered
sequence of variables near extremities. Such extremal quantities appear and
have determining effects in a variety of areas,  ranging from those in
physics like glasses \cite{BouchaudMezard:1997}, interface fluctuations and
random walks \cite{GyorgyiETAL:2003,LeDoussalMonthus:2003,MajumdarComtet:2004,SchehrMajumdar:2006},
front propagation \cite{KrapivskyMajumdar:2000} through engineering
\cite{Gumbel:1958} and  hydrology \cite{KatzParlangeNaveau:2002},
to seismology \cite{SornetteETAL:1996}, and finance
\cite{EmbrechtETAL:1997,Longin:2000}.  These and other applications make EVS
a remarkable field of multidisciplinary research, where theory and
data analysis meets practical challenges.

The asymptotic statistics of extreme values in sets of independent and
identically distributed (i.i.d.) variables has been known for a long
time \cite{FisherTippett:1928,Gnedenko:1943,Gumbel:1958}.  Three main
different types of limit distributions are traditionally distinguished
depending on the original distribution  (called parent distribution)
of the individual variables. If the density function of the parent
decays faster than any power law either at infinity or a finite upper border
then the asymptotic distribution is the Fisher-Tippett-Gumbel (FTG) function. If
the parent distribution has a power-law tail then  the limit distribution is
called Fisher-Tippett-Fr\'echet, and if it decays as a power function close to
an upper bound, it is the Fisher-Tippett-Weibull (FTW) distribution
\cite{Galambos:1978}.  These cases can be concatenated in a single,
one-parameter function form often called the generalized extreme value
distribution \cite{vonMises:1954,CluselBertin:2008}.

The question of corrections to the asymptotic distributions obviously arise in
applications, since practical problems involve finite data sets. Knowing the
magnitude and shape of the finite size corrections to the limit distribution is
thus of primary importance when analyzing EVS in empirical data. The
significance of finite size corrections has been recognized as early as in the
founding paper of the field \cite{FisherTippett:1928}. Then the warning was made
that logarithmically slow convergence may seriously hamper practical
applications of EVS \cite{Hall:1979}, and the first shape correction of
the limit distribution was calculated for a generalized Gaussian parent
\cite{Hall:1980}. The rigorous treatment of finite size corrections for the  EVS
of  i.i.d.\ variables can now be found in the mathematical literature
\cite{DeHaanResnick:1996,DeHaanStadm:1996}, but remains poorly known in the
physics community and in the multitude of other fields of application of EVS.

Recently a renormalization group (RG) approach was proposed to describe the
asymptotic behavior in EVS, worked out in detail for i.i.d.\ variables in
\cite{GyorgyiETAL:2008,GyorgyiETAL:2010}. Whereas it is widely known that the
concept of the RG has been the most efficient method for dealing
with critical phenomena \cite{ZinnJustin:2005}, its implementations have
been also useful for other problems in probability like for sums of variables,
i.e. the central limit problem area \cite{JonaLasinio:2001,CluselBertin:2008}.
In the field of EVS, an RG method was constructed to evaluate limit
distributions arising in random landscape and interface problems
\cite{LeDoussalMonthus:2003,SchehrLeDoussal:2009}.

The basic idea of RG methods in statistical physics is to perform some
coarse-graining by eliminating variables on one scale and defining effective
variables that describe the problem at some larger scale. Iterating this
procedure, in each step a finite operation, leads to the description of the
large scale behavior, possibly including singularities, of the system. In the
generic case, critical behavior is determined by a fixed point, while
near-critical physical systems are influenced by ``eigendirections'' of the
linearized RG transformation near the fixed point. Among other properties, the
finite size corrections to the infinite size behavior correspond to repelling,
i.e., relevant, eigendirections.  Concerning EVS, the landmark study of Fisher
and Tippett \cite{FisherTippett:1928} on the limit distributions can be
considered as a fixed point analysis of an appropriately defined RG operation.
This simple observation can be extended to the linear
neighborhood of fixed points, leading to the emergence of the finite size
correction functions \cite{GyorgyiETAL:2008,GyorgyiETAL:2010}. The difference to
the results in the mathematical  literature is that
\cite{DeHaanResnick:1996,DeHaanStadm:1996} obtained the rate of convergence and
shape corrections for families of initial parent distributions, while the RG
study, a priori, produces the shape correction functions from an eigenvalue
problem in function space, a technically quite simple proposition.  It turns out
\cite{GyorgyiETAL:2008,GyorgyiETAL:2010} that the index of eigenvalue is in fact
the exponent of the finite size correction, necessarily non-positive for a
vanishing correction. Thus the finite size corrections in EVS correspond to
stable perturbations about a fixed point, as opposed to the RG in
statistical physics based on the elimination of variables, where size
corresponds to unstable directions, but in agreement with the concept as
expounded  in probability theory by \cite{JonaLasinio:2001}. Naturally, also in
the RG treatment connection to the parent distribution should be established,
where the RG and the existing mathematical treatment meet, as done in
\cite{GyorgyiETAL:2008,GyorgyiETAL:2010}.

Given the manifest use of the RG approach in EVS, in this paper we reformulate
the RG study of \cite{GyorgyiETAL:2008,GyorgyiETAL:2010} in terms of a partial
differential equation (PDE) defining continuous flows in the space of
distributions.  In that way a simple and elegant method is obtained to recover the
known fixed points, lying on a fixed line, and the recently evaluated
eigendirections. This is a most manageable form of the RG transformation
proposed for EVS, possibly useful for generalizations like to higher
corrections, to other-than-extremal (order) statistics, or to correlated
variables. Interestingly, while the PDE is naturally thought as describing
smooth distributions, it is correctly handling discontinuities. In addition, the
PDE representation provides us with remarkable exact solutions to invariant
manifolds starting out from some unstable direction near the FTG fixed point. It
has been shown before in \cite{GyorgyiETAL:2010} that for some initial
distributions close to a fixed point, but differing from it in an unstable
eigenfunction, the distribution first goes away from the fixed point under
iteration of the RG transform, before either coming back to the same fixed point
or converging to another fixed point. Surprisingly, as it can be shown simply
via the present PDE approach, unstable perturbations near the FTG fixed point
conserve their functional form even while getting farther from the fixed point,
with only the amplitude and the finite-size-index depending on the flow
parameter, before the amplitude again becomes small during the final convergence
to the FTG limit. In other words, the invariant manifold starting out from and
returning to the FTG fixed point keeps being expressed in terms of a single
eigenfunction, albeit with a changing parameter. Paths in function space other
than the aforementioned ``excursions'' can also be found in terms of a single
eigenfunction, including those having a finite probabilistic mass at the upper
limit of the support. Even such distributions with  discontinuities are
described well by the PDE and are shown to converge to a Dirac delta, i.e.,
their limit distribution in EVS. These paths in distribution space, involving a
single eigenfunction with running eigenvalue index, can be generalized to
families of exactly calculated RG trajectories, where one starts out from a
continuously weighted combination of eigenfunctions about the FTG fixed point.
Remarkably, one can show that such is the form of the most general RG
trajectories in function space, irrespective whether they are near the fixed
line or away from it.  As the final area explored within the RG for EVS, we show
that exact, nonperturbative,  RG paths in terms of a single eigenfunction with
running parameters can also be given about FTW and FTF fixed points.
A seemingly paradoxical feature there is that a  single eigenfunction not
proper to the fixed point function can appear in those solutions.

Interestingly, if one thinks of the RG flow in more abstract terms, without
the motivation connecting it to EVS, it turns out that the present RG operation
is one of the simplest that can be defined, namely raising a function to a given
power, and then rescaling it. While in EVS this transformation is done on the
integrated distribution function, such an operation also appears in the well
known central limit problem \cite{Kolmogorov:1954,Christoph:1993} of sums
of random variables,  but there such an operation is performed on the moment
generating  (or characteristic) function.  This leads to a formal
similarity between EVS and the central limit problem, and so we can establish a
direct correspondence between Weibull distributions and symmetric
L\'evy-stable laws.

The paper is organized as follows.  In  Sec.\ \ref{S:basic} we motivate the
concept of RG in EVS. Section  \ref{S:rgflow} contains the main results, where
\ref{S:pde-rg-op} recalls basic notions including the definition of the RG operation, \ref{S:pde-rg} gives the derivation of the PDE describing the RG flow,
\ref{S:fp-pde} contains the fixed point condition, and
\ref{S:pert-fp-pde} presents the eigenvalue problem and its solutions about the
fixed point.  Section \ref{S:ustable} concerns the explicit, nonperturbative
solutions, wherein \ref{S:ustable-ftg} gives exactly invariant manifolds about
the FTG fixed point, the most general form  of exact RG paths is studied in
\ref{S:ustable-ftg2}, and \ref{S:ustable-ftf} contains the nonperturbative
solutions around FTW and FTF fixed points. Finally, Sec.\ \ref{S:clt} discusses
the formal similarity between EVS and the central limit theorem, focusing on
symmetric probability distributions in the latter case.  Appendix
\ref{S:ustable-ftg2-4} discusses equivalent forms of the most general RG path
via a direct method, without resorting to the differential formalism.

%%%%%%%%%%%%%%%%%%%%%%%%%%%%%%%%%%%%%%%%%%%%%%%%%%%%%%%%%%%%%%%%%%%%%%
%%%%%%
\section{Renormalization transformation for extreme values of i.i.d.\
random variables}
\label{S:basic}
%%%%%%%%%%%%%%%%%%%%%%%%%%%%%%%%%%%%%%%%%%%%%%%%%%%%%%%%%%%%%%%%%%%%%%
%%%%%%

In EVS the integrated distribution $\mu(x)$ is a most useful quantity,
defined for a random variable $x$ as
\be
\mu(x) = \int_{-\infty}^x \rho(x') dx',
\ee
where $\rho(x)$ is the probability density function.  It has the
meaning of the probability of finding the variable at any value below
$x$ -- apart from ambiguities due to Dirac deltas we do not treat here
separately.  In the simplest proposition of EVS, let us consider a set
of $N$ i.i.d.\ random variables $x_i$, $i=1,\ldots,N$.  The
probability that the maximum value in the set, $\max(x_1,\ldots,x_N)$,
is smaller than a given value $x$ is equal to the probability that all
the variables $x_i$ are less than $x$. As the variables are
statistically independent, this yields
\be
\mathrm{Prob}(\max(x_1,\ldots,x_N)<x) = \mu^N(x).
\ee
Often $\mu$ is called the parent distribution, whence the extreme
value distributions descend.

In what follows we give a flavor of the decimation method leading to
the RG formalism presented afterwards.  Let us split the set of
sufficiently large $N$ random variables $x_i$ into $N'=N/p$ blocks of
$p$ random variables each. Denoting by $y_j$ the maximum value in the
$j^{\mathrm{th}}$ block, one has
\be
\max(x_1,\ldots,x_N) = \max(y_1,\ldots,y_{N'}).
\ee
The variables $y_j$ are also i.i.d.\ random variables, with a
distribution
$\mu_p(y)$ given by
\be
\mu_p(y) = \mu^p(y). \label{eq:early-rg}
\ee
The above procedure, which can be further iterated, may thus be
thought of as a RG transformation, since a problem involving a large
number of random variables is transformed into a similar problem,
involving a reduced number of variables obeying a renormalized
distribution.  Note however that this is a quite simple example of RG,
as the variables are independent, while the RG was originally
designed,
in the context of critical phenomena, to deal with strongly correlated
variables.   Yet in EVS nonlinearities are important  even in the
i.i.d.\ case, and singularities may appear for large $N$, which can be
dealt with easily by the RG approach.   In particular, one can simply
derive universal shape corrections associated to finite size effects
that need a thorough mathematical study in a direct approach. Finally
we make a note of that the parameter $p$ is here a priori an integer
number $p\ge 2$.   In the following, we shall however consider the
case of a continuous variable.

%%%%%%%%%%%%%%%%%%%%%%%%%%%%%%%%%%%%%%%%%%%%%%%%%%%%%%%%%%%%%%%%%%%%%%
%%%%%%
\section{Renormalization flow}
\label{S:rgflow}
%%%%%%%%%%%%%%%%%%%%%%%%%%%%%%%%%%%%%%%%%%%%%%%%%%%%%%%%%%%%%%%%%%%%%%
%%%%%%

%%%%%%%%%%%%%%%%%%%%%%%%%%%%%%%%%%%%%%%%%%%%%%%%%%%%%%%%%%%%%%%%%%%%%%
%%%%%%
\subsection{The RG operation}
\label{S:pde-rg-op}
%%%%%%%%%%%%%%%%%%%%%%%%%%%%%%%%%%%%%%%%%%%%%%%%%%%%%%%%%%%%%%%%%%%%%%
%%%%%%

In previous papers \cite{GyorgyiETAL:2008,GyorgyiETAL:2010}, the RG
transformation has been introduced and studied, starting out from some parent
distribution and winding up near a universal limit function. While there
the iteration parameter $p$ was allowed to be continuous, when studying the fixed point and
the eigenfunctions about it, that parameter was discarded soon.  An alternative
approach, which as we shall see is well adapted to analytical studies, is to
consider the RG flow by means of a partial differential equation (PDE), obtained
from the discrete RG by continuation in $p$.  The RG transformation introduced
in \cite{GyorgyiETAL:2008,GyorgyiETAL:2010} had the form
\be
 [\hat{R}_p \mu]  (x)
= \mu^p\big(a(p)x+b(p)\big).
\label{eq:def-rg-former}
\ee
Here the scale and shift parameters $a(p),\, b(p)$, to be specified later, are
included in contrast to the native formula \eqref{eq:early-rg}.  They aim at
eliminating the degeneracy of the distribution developing in the large $p$ limit
and allow the emergence of a limit distribution.

As iterating $n$ times the RG transform with a scale factor $p$ corresponds to
an effective RG with a scale factor $p^n$, it is natural to introduce the
variable
\begin{equation}
  s=\ln p \label{eq:s-p}
\end{equation}
in order to parameterize the RG flow.  Indeed, iterations of the RG
flow then correspond to a linear increase in $s$.  The integrated
distribution after $p=e^s$ iterations will thus be denoted  as
$\mu(x,s)$. For later convenience, we also write the integrated
distribution into a double exponential form, namely
\be
\mu(x,s) =  e^{-e^{\scriptstyle -g(x,s)}},
\label{eq:def-g}
\ee
which defines the real function $g(x,s)$. In this way, $\mu(x,s)$ is necessarily
bounded between $0$ and $1$ for all real values of $g(x,s)$. The parent
distribution can be suitably taken as the function at $s=0$
\begin{subequations}
\begin{align}
\mu(x,0) &= \mu(x), \label{eq:ic1} \\
g(x,0) &= g(x), \label{eq:ic2}
\end{align}\label{eq:ic}
\end{subequations}
but we will also study trajectories in function space without reference to a
specific parent, when the origin of ``$s$'' will be set by some other
convention.

One of the major interests of the RG approach stems from the fact that it
facilitates the  study of the convergence to asymptotic distributions, including
the finite size corrections to this limit distributions.  However, from the
practical viewpoint limit distributions are equivalent up to a linear rescaling
$x \to ax+b$ of the random variable, with arbitrary finite parameters $a$ and
$b$. In order to lift this ambiguity, we impose the following standardization
conditions on the integrated distributions (two conditions are needed since
there are two free parameters $a$ and $b$ in the above rescaling):
\begin{subequations}
\begin{gather}
\mu(0,s) \equiv e^{-1}, \\
\partial_x \mu(0,s) \equiv e^{-1}.
\end{gather}
\label{eq:std-mu}
\end{subequations}
In terms of the function $g(x,s)$, these conditions translate into
\begin{subequations}
\begin{gather}
g(0,s) \equiv 0 \label{eq:std-ga}\\
\partial_x g(0,s) \equiv 1. \label{eq:std-gb}
\end{gather}
\label{eq:std-g}
\end{subequations}
This convention of standardization will be kept throughout the paper.

%%%%%%%%%%%%%%%%%%%%%%%%%%%%%%%%%%%%%%%%%%%%%%%%%%%%%%%%%%%%%%%%%%%%%%
%%%%%%
\subsection{Partial differential equation of the flow}
\label{S:pde-rg}
%%%%%%%%%%%%%%%%%%%%%%%%%%%%%%%%%%%%%%%%%%%%%%%%%%%%%%%%%%%%%%%%%%%%%%
%%%%%%

%%%%%%%%%%%%%%%%%%%%%%%%%%%%%%%%%%%%%%%%%%%%%%%%%%%%%%%%%%%%%%%%%%%%%%
\subsubsection{RG transformation in the second exponent}
\label{S:pde-rg-second-exp}

We now turn to the derivation of the PDE for the RG flow. Firstly, with the
reparameterization \eqref{eq:s-p} we convert the subscript and argument from $p$
to $s$ in the RG transform \eqref{eq:def-rg-former}, yielding the renormalized
integrated distribution as
\be \mu(x,s) \equiv [ \hat{R}_{s} \mu ] (x)
= \mu^{e^s}\! \big(a(s)x+b(s)\big),
\label{eq:def-rg}
\ee wherein the functions $a(s)$ and $b(s)$  enforce the standardization
conditions \eqref{eq:std-mu}. In terms of the function $g(x,s)$, the RG
transform \eqref{eq:def-rg} can be rewritten, by taking the double logarithm, as
\be
g(x,s) = g\big(a(s)x+b(s)\big) - s.\label{eq:gxs}
\ee
This is a very simple operation, a linear change of variable in the argument
of the parent \eqref{eq:ic2} and a global shift, wherein, however,  $a(s)$ and
$b(s)$ are determined  so as to satisfy the nonlinear standardization
conditions \eqref{eq:std-g}.

%%%%%%%%%%%%%%%%%%%%%%%%%%%%%%%%%%%%%%%%%%%%%%%%%%%%%%%%%%%%%%%%%%%%%%
\subsubsection{Driven linear PDE}
\label{S:pde-rg-pde-driven}
%%%%%%%%%%%%%%%%%%%%%%%%%%%%%%%%%%%%%%%%%%%%%%%%%%%%%%%%%%%%%%%%%%%%%%

In order to make the RG transformation more explicit we now compute the
parameters $a(s)$ and $b(s)$. Setting $x=0$ in \eqref{eq:gxs}, we get from
\eqref{eq:std-ga} that $g(b(s))=s$,
so that
\be
b(s) = g^{-1}(s). \label{eq:bs}
\ee
Note that $g(x)$ is a monotonously increasing function,
invertible over the support of the parent distribution.
To determine $a(s)$, we first differentiate \eqref{eq:gxs} with
respect to $x$, yielding
\be
\partial_x g(x,s) =a(s) g'\big(a(s)x+b(s)\big).\label{eq:gxsx}
\ee
Setting again $x=0$ and using Eq.~\eqref{eq:std-gb}, we get
\be
a(s)= 1/g'(b(s)).\label{eq:as}
\ee

Since we aim at describing the continuous evolution of $g(x,s)$ as a
function of $s$, we differentiate \eqref{eq:gxs} with respect
to $s$ and obtain
\be
\partial_s g(x,s) = \big(\dot{a}(s) x + \dot{b}(s)\big)
g'\big(a(s)x+b(s)\big) - 1,
\label{eq:gxss}
\ee
where the dot denotes the derivative with respect to $s$.
Combining equations \eqref{eq:bs} and \eqref{eq:as}, we get
\begin{equation}
\dot{b}(s)=a(s), \label{eq:a-bdot}
\end{equation}
which we substitute into Eq.~\eqref{eq:gxss}.  Then we use
\eqref{eq:gxsx} in order to eliminate $g'$ from Eq.~\eqref{eq:gxss},
and if we introduce as well the notation
\begin{equation}
\gamma(s)= \frac{\dot{a}(s)}{a(s)} = - \frac{g''(b(s))}{g'^2(b(s))},
\label{eq:gamma-s}
\end{equation}
we eventually obtain the PDE
\begin{equation}
\partial_s g(x,s) = (1+\gamma(s)x) \partial_x g(x,s) -
1.\label{eq:pde}
\end{equation}
This equation should be taken with initial condition defined by the parent
distribution through \eqref{eq:ic}, and the parent also determines $\gamma(s)$
by \eqref{eq:gamma-s}.  We emphasize that the original form \eqref{eq:def-rg} of
the RG transformation produces valid (monotonically increasing in $x$)
distribution functions from a like initial function, thus the PDE, derived from
\eqref{eq:def-rg}, also must preserve monotonicity.   This is a property
maintained even if it is not obvious directly from the PDE. We note furthermore,
that if the standardization condition \eqref{eq:std-ga} is met then, as it is
easy to see, the solution of the PDE automatically satisfies \eqref{eq:std-gb}.

In \eqref{eq:pde} the coefficient function $\gamma(s)$ can be considered as an
``external'' forcing, which, beside the initial condition, carries information
on the parent. In this sense, \eqref{eq:pde} is a linear, parametrically driven
PDE.

%%%%%%%%%%%%%%%%%%%%%%%%%%%%%%%%%%%%%%%%%%%%%%%%%%%%%%%%%%%%%%%%%%%%%%
%%%%%%
\subsubsection{Autonomous, nonlinear PDE}
\label{S:pde-rg-pde-aut}
%%%%%%%%%%%%%%%%%%%%%%%%%%%%%%%%%%%%%%%%%%%%%%%%%%%%%%%%%%%%%%%%%%%%%%
%%%%%%

Interestingly, the same PDE \eqref{eq:pde} can be alternatively interpreted as
an  autonomous but nonlinear flow equation.  For this purpose we start out
from an arbitrary point on the flow $g(x,s)$ and apply an infinitesimal
renormalization transformation to it as
\be
g(x,s+ds)=[\hat{R}_{ds}g](x,s).
\ee
Consistently with \eqref{eq:gxs}, this involves a linear change of variable and
an overall shift, both infinitesimal as
\be g(x,s+ds) = g\Big(\big(1+\tilde a(s)ds\big)x+\tilde b(s) ds,s\Big) -ds,
\label{eq:RGauton1}
\ee
wherein  the functions $\tilde a(s)$ and $\tilde b(s)$ are to be specified.
Linearizing \eqref{eq:RGauton1} with respect to $ds$, we get
\be
\partial_s g(x,s) =
\big(\tilde a(s)x+\tilde b(s)\big)\partial_x g(x,s) -1.
\label{eq:RGauton2}
\ee
Using the standardization conditions \eqref{eq:std-g} we can hence determine
the unknown functions $\tilde a(s)$ and $\tilde b(s)$. Setting $x=0$ yields a
constant $\tilde b(s)\equiv1$, whence we recover the previously derived PDE
\eqref{eq:pde}, where in place of $\tilde a(s)$ the $\gamma(s)$ appears, the
symbol we shall use for the coefficient function henceforth.  As a final step in
this reasoning, by once differentiating \eqref{eq:pde} in terms of $x$, setting
$x=0$, and using \eqref{eq:std-gb} we find that
\begin{equation} \gamma(s) = - \partial_x^2 g(0,s)\label{eq:gammas}
\end{equation}
must hold.   So in this picture the PDE is the same as \eqref{eq:pde},
wherein the coefficient function $\gamma(s)$ is the initial curvature
\eqref{eq:gammas} of the field $g(x,s)$. Then one can try to find families of
solutions of the PDE, where $\gamma(s)$ is calculated from the
self-consistency condition \eqref{eq:gammas}. Note that the presence
of $\gamma(s)$ defined by \eqref{eq:gammas} makes the PDE nonlinear, and
actually nonlocal because the curvature at $x=0$ affects the evolution at all
$x$'s, but without external driving.

%%%%%%%%%%%%%%%%%%%%%%%%%%%%%%%%%%%%%%%%%%%%%%%%%%%%%%%%%%%%%%%%%%%%%%
\subsubsection{Dual interpretation}
\label{S:pde-rg-pde-sum}
%%%%%%%%%%%%%%%%%%%%%%%%%%%%%%%%%%%%%%%%%%%%%%%%%%%%%%%%%%%%%%%%%%%%%%

In conclusion, the RG flow is described by \eqref{eq:pde}, where the
coefficient function $\gamma(s)$  is  either an external forcing given by the
parent $g(x)$ as described in paragraph \ref{S:pde-rg-pde-driven} or defined by
the initial curvature of the field $g(x,s)$ itself as in
\ref{S:pde-rg-pde-aut}.   The two definitions \eqref{eq:gamma-s} and
\eqref{eq:gammas} are, however, identical, as one can show by using the
standardization \eqref{eq:std-g} and the form \eqref{eq:gxs} for the RG
transformation.   Thus the PDE representation of the RG transformation has  a
``dual`` interpretation, which can be thought of as a consequence of the
stringent standardization requirements. The PDE \eqref{eq:pde} is the central
result of the present paper and the source of the subsequent analysis.

%%%%%%%%%%%%%%%%%%%%%%%%%%%%%%%%%%%%%%%%%%%%%%%%%%%%%%%%%%%%%%%%%%%%%%
%%%%%%
\subsubsection{Discontinuities in the
integrated distribution }
\label{S:pde-rg-supp}
%%%%%%%%%%%%%%%%%%%%%%%%%%%%%%%%%%%%%%%%%%%%%%%%%%%%%%%%%%%%%%%%%%%%%%
%%%%%%

An inherent property of the distribution is the extent of its support, i.e.\ the
(closure of the) range in $x$ where the integrated distribution is strictly
monotonically increasing. Outside of the support the probability densities vanish.

Firstly we mention that limited supports arise in applications and even the
known limit distributions of EVS have  limited support in the FTW and FTF cases
\cite{FisherTippett:1928}.  We emphasize, however, that the representation
\eqref{eq:def-g} allows us to handle the case when the support is limited to
some range $[x_\ast(s),x^\ast(s)]$ of $x$ wherein the integrated distribution
$\mu(x,s)$ smoothly increases from $0$ to $1$.  Indeed there the function
$g(x,s)$ increases smoothly from $-\infty$ to $\infty$ and  no obstacle seems to
arise before the application of the PDE \eqref{eq:pde}.

A more problematic situation emerges when the distribution has a ``point
measure'', i.e.\ the density has a Dirac delta somewhere.  This means a
discontinuity in the integrated distribution and thus in the $g$ function,
making the PDE ill-defined at first sight.  Given the fact that later in the
paper we will encounter such distributions here we show that such cases can be
treated consistently within the PDE formalism as well.

Note firstly that if the distribution $\mu(x,0)=\mu(x)$ has a discontinuity at
$x_{\mathrm{dsc}}$ then the RG transformation \eqref{eq:def-rg}, for some other
parameter setting $s$, will result in another distribution $\mu(x,s)$ exhibiting
a discontinuity at the transformed $x_{\mathrm{dsc}}(s)$ satisfying
\begin{equation}
x_{\mathrm{dsc}} = a(s) \, x_{\mathrm{dsc}}(s) + b(s).
\label{eq:discontinuity1}
\end{equation}
After differentiating and using Eqs.\ (\ref{eq:a-bdot},\ref{eq:gamma-s}) we get
\begin{equation}
\dot x_{\mathrm{dsc}}(s) = - 1 - \gamma(s)\,x_{\mathrm{dsc}}(s).
\label{eq:discontinuity}
\end{equation}
This equation determines the evolution of the discontinuity point in $s$ under
the original form of the RG transformation \eqref{eq:def-rg}.

Now we shall consider the PDE representation of the RG \eqref{eq:pde} in the
neighborhood of the discontinuity.  Assuming a finite discontinuity in $g$ of
step size $D$ we have approximately
\begin{equation}
 g(x,s)\approx D\, \theta(x-x_{\mathrm{dsc}}(s)) + E,
\label{eq:g-theta}
\end{equation}
where $D,E$ are constants and $\theta(x)$ denotes the Heaviside symbol.
Substituting that into \eqref{eq:pde} we obtain
\begin{equation}
\begin{split}
  - D\, \dot x_{\mathrm{dsc}}(s) &\,\delta(x-x_{\mathrm{dsc}}(s)) \\
&\approx
(1+\gamma(s)\, x)\, D\, \delta(x-x_{\mathrm{dsc}}(s))  - 1,
\label{eq:pde-disc}
\end{split}
\end{equation}
with the Dirac delta notation $\theta'\!(x)=\delta(x)$.  Equating the
coefficients of the Dirac deltas we recover Eq.\ \eqref{eq:discontinuity}.

The above consideration can be extended to the case when the discontinuity is at
one of the lower or upper borders of the support.  There the integrated
distribution jumps from zero, or to one, which corresponds by \eqref{eq:def-g}
to jumps from $-\infty$, or to $\infty$, in the function $g(x,s)$, respectively.
The infinite jumps can be conceived as \eqref{eq:g-theta} with diverging step
size $D$, yielding in the end the same condition \eqref{eq:discontinuity}.

In conclusion we reemphasize that, if the initial distribution contains a
discontinuity, this will move just as prescribed by the linear change of
variable in the RG transformation.   Moreover, the singularity analysis of the
PDE consistently yields the equation for the moving discontinuity, so the PDE
representation is upheld even for non-smooth distributions.  The consistency in
the evolution of the size of the jump needs further study, but here we
concentrated on its location since this will be discussed on examples later in
the paper.
%%%%%%%%%%%%%%%%%%%%%%%%%%%%%%%%%%%%%%%%%%%%%%%%%%%%%%%%%%%%%%%%%%%%%%
%%%%%%
\subsection{Fixed point}
\label{S:fp-pde}
%%%%%%%%%%%%%%%%%%%%%%%%%%%%%%%%%%%%%%%%%%%%%%%%%%%%%%%%%%%%%%%%%%%%%%
%%%%%%

Let us first look for stationary solutions  $g(x,s)\equiv f(x)$ of the PDE
\eqref{eq:pde} independent of $s$.  Then $f(x)$ is in fact a fixed point and
must satisfy the stationary version of \eqref{eq:pde}
\begin{equation}
0 = (1+\gamma x) f'(x) - 1,\label{eq:pde-fpeq}
\end{equation}
where by \eqref{eq:gammas} we must have a constant $\gamma$ arising in
\begin{equation}
 \gamma = - f''(0). \label{eq:gammaf}
\end{equation}
Hence, using the standardization condition $f(0)=0$, we get
\begin{equation}
f(x;\gamma) = \int_0^x (1+\gamma y)^{-1}dy = \frac 1\gamma
\ln(1+\gamma x) ,\label{eq:pde-fp}
\end{equation}
whence the fixed point distribution is
\begin{equation}
 M(x;\gamma)= e^{-e^{\scriptstyle  -f(x;\gamma)}} = e^{-(1+\gamma
x)^{-1/\gamma}}.
\end{equation}
This is the well known generalized extreme value distribution, obtained here as
a fixed line of the RG transformation \cite{FisherTippett:1928,Gumbel:1958}. The
above derivation of the universal limit distribution family is quite brief, and
was made possible by the PDE-representation of the RG flow.

%%%%%%%%%%%%%%%%%%%%%%%%%%%%%%%%%%%%%%%%%%%%%%%%%%%%%%%%%%%%%%%%%%%%%%
%%%%%%
\subsection{Perturbations about a fixed point}
\label{S:pert-fp-pde}
%%%%%%%%%%%%%%%%%%%%%%%%%%%%%%%%%%%%%%%%%%%%%%%%%%%%%%%%%%%%%%%%%%%%%%
%%%%%%

\subsubsection{RG flow for arbitrary deviations}

In order to study the behavior of the flow off the fixed line, we perturb a
fixed point function \eqref{eq:pde-fp} by some $\phi(x,s)$, not necessarily
small, and then write for it the flow equation based on (\ref{eq:pde}). We
consider two forms of perturbed distribution,  the first choice being a modified
argument as
%\begin{subequations}
% \begin{align}
% g(x,s) &= f(x+\phi(x,s))  \label{eq:g-psi1}\\
% g(x,s) &=  f(x)+f'(x)\phi(x,s) \label{eq:g-psi2}
% \end{align}
% \label{eq:g-psi2}
% \end{subequations}
% \!\!\!\!\!\!
\begin{equation}
 g(x,s) = f\big(x+\phi(x,s)\big).  \label{eq:g-psi1}
\end{equation}
For now  the argument $\gamma$ has been omitted and  $f(x)\equiv f(x;\gamma)$
used to simplify notation.  It is easy to see that \be \label{eq:std-phi}
\phi(0,s)=0, \quad \partial_x \phi(0,s)=0 \ee satisfies the standardization
condition \eqref{eq:std-g}. Thus, using \eqref{eq:pde-fp}, we have for the
$\gamma(s)$ of Eq.~(\ref{eq:gammas})
\be
\gamma(s) =\gamma -\partial_x^2 \phi(0,s).\label{eq:gammas2}
\ee
Eventually,  by substituting Eq.~\eqref{eq:g-psi1} into (\ref{eq:pde}) and using
\eqref{eq:pde-fp}, after obvious manipulations we obtain the PDE for the
perturbation function $\phi(x,s)$
\be \label{eq:lin-pde1}
\begin{split}
\partial_s \phi(x,s) &=  (1+\gamma(s) x)\,  \partial_x \phi(x,s) -
\gamma\, \phi(x,s) \\&\quad + (\gamma(s) - \gamma)\, x,
\end{split}
\ee
This is the form of perturbation we shall use in what follows.
However, as an interesting side remark, we note that the way to introduce
the perturbation is not unique, as soon as calculations are considered beyond
linear order in the perturbation.
Indeed, another standard possibility is to modify
the fixed point function additively, in the form
\begin{equation}
 g(x,s) = f(x)+f'(x)\,\phi(x,s),  \label{eq:g-psi2}
\end{equation}
so that to linear order in $\phi(x,s)$ it be equivalent to \eqref{eq:g-psi1}.
This ansatz implies again the relations (\ref{eq:std-phi},\ref{eq:gammas2}). The
PDE for the thus defined $\phi(x,s)$ is then obtained by substituting
Eq.~\eqref{eq:g-psi2} into (\ref{eq:pde})
\begin{align}
\partial_s \phi(x,s) &=  (1+\gamma(s) x)  \big( \partial_x \phi(x,s) -
\gamma f'(x)\, \phi(x,s)\big) \nonumber \\ &\quad + (\gamma(s) -
\gamma) \, x. \label{eq:lin-pde2} \end{align}
We emphasize that the RG flow equations (\ref{eq:lin-pde1},\ref{eq:lin-pde2})
are exact, corresponding to the definitions (\ref{eq:g-psi1},\ref{eq:g-psi2}),
respectively.  Note that the two PDE's coincide only, if $\gamma(s)= \gamma$ or
$\gamma=0$. In the latter case
\be \label{eq:lin-pde-ftg}
\partial_s \phi(x,s) =  (1+\gamma(s) x)\, \partial_x \phi(x,s)
+\gamma(s)\, x,
\ee
which we display for later usage.

\subsubsection{Linear perturbations about a fixed point}

Since the above two definitions of the perturbations are equivalent to linear
order in $\phi(x,s)$, the corresponding PDE's
(\ref{eq:lin-pde1},\ref{eq:lin-pde2}) should become identical after
linearization. From \eqref{eq:gammas2} we see that the smallness of $\phi(x,s)$
implies that $\gamma(s)$ is close to $\gamma$.  After linearization of either
PDE (\ref{eq:lin-pde1},\ref{eq:lin-pde2})  we wind up with
\be
\label{eq:lin-pde3}
\begin{split}
\partial_s \phi(x,s) &=  (1+\gamma x)  \,\partial_x \phi(x,s) -
\gamma\, \phi(x,s)  \\ &\quad - x\,\partial^2_x\phi(0,s).
\end{split}
\ee
In what follows we determine the eigenfunctions emerging from this PDE and also
discuss its general solution.

\subsubsection{Eigenfunctions}

Here we are seeking the eigenfunctions of Eq.~(\ref{eq:lin-pde3}) defined by the
property that they evolve by a purely exponential $s$-dependence. As we shall
see, these are precisely those functions $\phi(x,s)$ in which the $s$ and $x$
dependence are factorized as
\begin{equation}
 \phi(x,s)= \epsilon(s)\, \psi(x). \label{eq:g-psi}
\end{equation}
Then the standardization \eqref{eq:std-phi} implies
\begin{subequations}
\begin{align}
 \psi(0)&=0, \\
 \psi'(0)&=0.
\end{align} \label{eq:stand-psi}
\end{subequations}
The factorization \eqref{eq:g-psi} is however not unique, as multiplying
$\epsilon(s)$ by a constant and dividing $\psi(x)$ by the same constant yields
the same $\phi(x,s)$. This ambiguity can be eliminated by setting the scale of
$\psi(x)$ through an additional condition; we choose
\begin{align}
\psi''(0)&=-1. \label{eq:stand-psi2}
\end{align}
Then condition \eqref{eq:gammas2} straightforwardly gives
\begin{equation}
\epsilon(s) = \gamma(s) - \gamma. \label{eq:eps-def}
\end{equation}
This simple relation justifies the somewhat unusual condition
\eqref{eq:stand-psi2}; furthermore, obviously $\epsilon(s)$ must be small for
the linearization remaining valid. Inserting \eqref{eq:g-psi} into the
linearized PDE \eqref{eq:lin-pde3} we get
\begin{equation}
\begin{split}  \label{eq:pde-eps}
\dot{\epsilon}(s) \psi(x) &= \epsilon(s)\big( (1+\gamma x ) \psi'(x)
 - \gamma  \psi(x) + x\big).
\end{split}
\end{equation}
This can be solved only if
\begin{equation}
 \frac{\dot{\epsilon}(s)}{{\epsilon}(s)} = \gamma'
\end{equation}
is a constant, whence
\begin{equation}
  \epsilon(s)\propto e^{\gamma' s} \label{eq:eps-exp}
\end{equation}and so
\begin{equation}
 (1+\gamma x) \psi'(x) = (\gamma+\gamma') \psi(x) - x.
\end{equation}
The homogeneous part has the solution
\begin{gather}
 \psi_H(x) \propto (1+\gamma x) ^{(\gamma+\gamma')/\gamma},
\end{gather}
whereas a particular solution of the inhomogeneous equation is the linear
function
 \begin{gather}
 \psi_P(x) = \frac{1+ (\gamma'+\gamma)x}{\gamma'(\gamma'+\gamma)}.
\end{gather}
The combination of the two so as to satisfy $\psi(0)=0$ necessarily
leads to the solution
\begin{equation}
\psi(x;\gamma,\gamma') = \frac{1+ (\gamma'+\gamma)x   -
(1+\gamma x)^{\gamma'/\gamma+1}}
{\gamma'(\gamma'+\gamma)},
\label{eq:psi-rg}
\end{equation}
where we reinserted the parameter arguments into $\psi$. Note that the condition
$\psi'(0)=0$ is also satisfied, as it should. The function
$\psi(x;\gamma,\gamma')$ is defined for $x$ in the range $1+\gamma x>0$. In the
case $\gamma=0$, the function $\psi(x;0,\gamma')$ becomes
\be
\psi(x;\gamma') = \frac{1}{\gamma'^2} \left(1+\gamma' x-e^{\gamma'
x}\right).
\label{eq:psi-FTG}
\ee
The eigenfunction \eqref{eq:psi-rg} is the same as obtained in
\cite{GyorgyiETAL:2008,GyorgyiETAL:2010} not using PDE.  There also the
empirical meaning of the above results has been clarified: interpreting $N=e^s$
as the number of variables, a fixed point limit distribution is assumed to be
reached for $N\to\infty$ as $g(\ln N,x)\to f(x)$, and then $\gamma'\le 0$ is the
exponent of decay in $N$ of the correction to the limit distribution. Note that
this interpretation slightly differs from the one presented in Sec.\
\ref{S:basic}, where $p=e^s$ was understood, and we kept $N$ finite while
performing decimation.

\subsubsection{General solution for linear perturbations}
\label{S:gen-lin-sol}

We now turn to more general solutions of the linearized Eq.~(\ref{eq:lin-pde3}).
Starting from the observation that a linear superposition of different
eigenfunctions is also a solution, after some considerations we arrive at
\be\label{sol-gen-eqlin}
\begin{split}
\phi(x,s) &= (1+\gamma x) \,\chi(s) + x\chi'(s)\\
&  \qquad - (1+\gamma x)\, \chi(f(x)+s),
\end{split}
\ee
in terms of an appropriate, single argument  $\chi(x)$ function.  Such a
solution satisfies by construction the standardization conditions
(\ref{eq:std-phi}).  We propose that the above form is the general solution of
the linearized RG flow equation (\ref{eq:lin-pde3}).

By the specific choice
\be
\chi(s) \propto e^{\gamma' s}
\ee
we recover the solution $\phi$ proportional to the factorized form
\eqref{eq:g-psi} with \eqref{eq:eps-exp} and the eigenfunction
\eqref{eq:psi-rg}.  Choosing now $\chi(s)$ as a linear combination of
such
exponential functions we obtain
\be \label{hs-lin-exp}
\chi(s) = \int d\gamma' p(\gamma')\, e^{\gamma' s},
\ee
where $p(\gamma')$ is a weight function,  rendering the above integral finite.
The formula for $\chi(s)$ is reminiscent to the Laplace transform of the weight
function $p(\gamma')$.   Obviously, if positive $\gamma'$ indices are allowed by
the weight function $p(\gamma')$ then $\chi(s)$  will grow in $s$, so the
solution $\phi(x,s)$ remains perturbative only for a limited range in $s$ and
$x$.  Note that we do not study the question of support, i.e., the range in $x$
where the linear solution corresponds to a valid, increasing distribution, but
understand that if the perturbation is small then the $g(x,s)$ of Eq.\
\eqref{eq:g-psi1} will be increasing in some range of $x$ about the origin.

%%%%%%%%%%%%%%%%%%%%%%%%%%%%%%%%%%%%%%%%%%%%%%%%%%%%%%%%%%%%%%%%%%%%%%
%%%%%%
\section{Exact nonperturbative RG trajectories}
\label{S:ustable}
%%%%%%%%%%%%%%%%%%%%%%%%%%%%%%%%%%%%%%%%%%%%%%%%%%%%%%%%%%%%%%%%%%%%%%
%%%%%%

An interesting phenomenon, already recognized in \cite{GyorgyiETAL:2010} is that
iterating the RG transform starting from a distribution close to the FTG
distribution can lead to an excursion in function space quite far from the FTG
distribution, before eventually coming back to it. In other words, the FTG fixed
point has both stable and unstable directions, and unstable manifolds generated
by unstable directions loop back along stable directions towards the same fixed
point.  Concerning the FTW and FTF fixed points, with $\gamma\neq 0$, a
trajectory starting out from it also returns to a fixed point, but the latter
may or may not be the same as the starting one.

Remarkably, by means of the PDE formalism one can show that the aforementioned
RG trajectory, starting near the FTG fixed point with an initial perturbation
along a single unstable eigenfunction of the FTG class remains within the
subspace of the eigenfunctions all along the trajectory, even in the non-linear
regime, before it eventually returns to the same fixed point. Actually, this
looping solution is one of several different families of those special RG paths,
whose full trajectories are specified by a single eigenfunction with running
perturbation $\epsilon$ and eigenvalue parameter $\gamma'$.

Subsequently we consider nonperturbative solutions obtained from
a reparameterization of the general solution of the linearized equation about
the FTG function. It
will turn out that this case in fact represents all general RG trajectories with
arbitrary parent distribution, i.e., initial condition.
We also clarify the relationship between this reparameterization and the
basic RG transform, which may also be thought of as a reparameterization
of the parent distribution.

Finally, further trial functions for trajectories  involving a single
eigenfunction about a general fixed point with $\gamma\neq 0$, containing
running parameters  $\epsilon$, $\gamma$, and $\gamma'$ will also be studied and
examples given when they represent exact RG flows.

\subsection{RG flow in terms of a single eigenfunction about FTG}
\label{S:ustable-ftg}

\subsubsection{The distribution and its support}
\label{S:ansaty-ftg1}

We shall now proceed to show how explicit RG trajectories emerge, where the
deviation from the FTG fixed point can be given in terms of a single
eigenfunction with an appropriate parameterization.  That way the RG
transformation in distribution space reduces to a two-dimensional flow for the
perturbation strength $\epsilon$ and the eigenvalue parameter $\gamma'$, which
we now set out to determine.

Let us consider the ansatz for the function $g(x,s)$, deviating from the FTG
fixed point $f(x;0)= x$, as
\be
g(x,s) = x + \epsilon(s) \psi\big(x;\gamma'(s)\big),
\label{eq:FTG-ansatz}
\ee
where the eigenfunction $\psi(x;\gamma')$ is given by \eqref{eq:psi-FTG}. Here
the function $g(x,s)$ depends on $s$ only through the functions $\epsilon(s)$
and $\gamma'(s)$, to be specified in the next subsection. We remind the reader
that the function $g(x,s)$ determines through \eqref{eq:def-g} the distribution,
so if $\epsilon(s)$  is not small then the distribution will depend on it
nonlinearly.

In order to fully specify the distribution, also its support needs to be given.
The function $g(x,s)$ as defined in \eqref{eq:def-g} was supposed to be
monotonically increasing in $x$ from $-\infty$  to $+\infty$ over the support,
however, the above function \eqref{eq:FTG-ansatz} may violate this condition.
Inspection of \eqref{eq:FTG-ansatz} shows that one extremum can develop at
\begin{equation}
 x_{\mathrm{extr}} = \frac{1}{\gamma'} \ln \left( 1+
\frac{\gamma'}{\epsilon}\right), \label{eq:xextr}
\end{equation}
if the argument of the logarithm is positive.  The $ x_{\mathrm{extr}}$ is a
maximum, if $\epsilon > \max(-\gamma',0)$ and it is a minimum, if $\epsilon <
\min(-\gamma',0)$.   In both cases there is a semi-infinite range in $x$ where
$g(x,s)$ decreases, which we exclude and define the entire region of monotonic
increase as the support.  In particular, denoting by  $x_\ast$ and $x^\ast$ the
lower and upper border of the support, respectively, we obtain
\begin{subequations}
\begin{align}
x_\ast &=\left\lbrace
\begin{array}{lr}
-\infty ,\ \mathrm{if} & \epsilon > \min(-\gamma',0), \\
x_{\mathrm{extr}} , \ \mathrm{if}&   \epsilon < \min(-\gamma',0),\\
\end{array}\right. \\
x^\ast &= \left\lbrace
\begin{array}{lr}
\infty , \ \mathrm{if} & \epsilon < \max(-\gamma',0), \\
x_{\mathrm{extr}} , \ \mathrm{if} &   \epsilon > \max(-\gamma',0).\\
\end{array}\right.
\end{align} \label{eq:borders}
\end{subequations}
\!\!\!\!  Note that $x_{\mathrm{extr}}$ depends on $s$ through the parameters
$\epsilon(s)$ and $\gamma'(s)$.  In the excluded region we simply define the
integrated distribution $\mu(x,s)$ as constant
\begin{equation}
\mu(x,s)  \equiv \left\lbrace
\begin{array}{lr}
0 , \ \mathrm{if} & x<x_\ast, \\
1 , \ \mathrm{if} &  x>x^\ast,\\
\end{array}\right.\label{eq:mu-const}
\end{equation}
and, equivalently, by \eqref{eq:def-g} we must have
\begin{equation}
g(x,s)  \equiv \left\lbrace
\begin{array}{rr}
-\infty  , \  \mathrm{if} & x<x_\ast, \\
\infty , \  \mathrm{if} &  x>x^\ast.\\
\end{array}\right.\label{eq:g-const}
\end{equation}
By this truncation we arrive at a valid, monotonically increasing
distribution.

The connection to the PDE is upheld such  that it determines the evolution of
the monotonically increasing part of $g(x,s)$.  Then the evolution of the
possibly finite extremum point $x_{\mathrm{extr}}(s)$ is also given, and there
we simply understand in $g(x,s)$ the jump from $-\infty$ to
$g(x_{\mathrm{extr}}(s),s)$ if it is minimum, or, if it is a maximum, from
$g(x_{\mathrm{extr}}(s),s)$ to $+\infty$.  In order to maintain consistency with
the PDE, as it has been discussed in Sec.\ \ref{S:pde-rg-supp}, such a
discontinuity point has to satisfy \eqref{eq:discontinuity}, a condition we
check later.

Finally we mention that not included in formula \eqref{eq:borders} are
$\epsilon=0$, which is the FTG case with support being the real axis, and
$\epsilon=-\gamma'$, where $g(x,s)$ is monotonically increasing over the real
axis, but does not go from $-\infty$  to $+\infty$ as needed by \eqref{eq:def-g}
to give a valid distribution. In other words, in the latter case the density is
not normalized to unity. Here a finite limit of the support and the necessary
jump in $g(x,s)$ can also be introduced consistently with
\eqref{eq:discontinuity}, so that way, while there is an arbitrariness in
truncating the support, a valid distribution can be associated also to this
case.

\subsubsection{Flow equations and their solution}
\label{S:ode-ftg1}

We know that for $|\epsilon|$ small and $\gamma'>0$ the function
(\ref{eq:FTG-ansatz})  represents an unstable direction near the FTG fixed
point, thus $\epsilon(s)\propto e^{s\gamma'}$ so long as it can be considered
small. Now we determine under what conditions the function form
(\ref{eq:FTG-ansatz}) remains valid even beyond the region linear in $\epsilon$.
 This ansatz is a special case of either (\ref{eq:g-psi1}) or (\ref{eq:g-psi2}),
identical for  $\gamma=0$, and with
\begin{equation}
\phi(x,s) =  \epsilon(s) \psi(x;\gamma'(s)).
\end{equation}
The definition \eqref{eq:gammas2}, together with the standardization
\eqref{eq:stand-psi2}, now specializes to
\be
\gamma(s) = \epsilon(s).
\ee
Substituting the above $\phi(x,s)$ and $\gamma(s)$ into the evolution
equation \eqref{eq:lin-pde-ftg}  we arrive at
\be
\dot{\epsilon}\,\psi + \epsilon\,\dot{\gamma}'\ \partial_{\gamma'}\psi
- (1 + \epsilon x)\, \epsilon\, \partial_x \psi - \epsilon\, x = 0.
\label{eq:to-ode1}
\ee
From the expression (\ref{eq:psi-FTG}) of $\psi(x;\gamma')$, we obtain
\begin{subequations}
\begin{align}
\partial_{\gamma'} \psi &= -\frac{2}{\gamma'} \psi +
\frac{x}{\gamma'}\partial_x \psi \\
\partial_x \psi &=\gamma' \psi-x.
\end{align}
\end{subequations}
We thus end up with
\be
\left( \dot{\epsilon}-2\epsilon\frac{\dot{\gamma}'}{\gamma'}
- \epsilon\gamma' \right) \psi
+ \left( \epsilon\frac{\dot{\gamma}'}{\gamma'} - \epsilon^2 \right)
x\partial_x
\psi =0.  \label{eq:to-ode2}
\ee
As the functions $\psi$ and $x\partial_x \psi$ are linearly independent, the
expressions between brackets in the above equation must vanish, leading to two
coupled evolution equations for $\epsilon$ and $\gamma'$. After rearrangement we
wind up with
\begin{subequations} \label{eq:eps-gammap-ode}
\begin{align}
\dot{\epsilon} &= 2\epsilon^2 + \epsilon \gamma', \label{eq:eps}\\
\dot{\gamma}' &= \epsilon\gamma'. \label{eq:gammap}
\end{align}
\label{eq:eps-gamma}
\end{subequations}
\!\!\!\! Thus the RG path for the ansatz \eqref{eq:FTG-ansatz} has been led back to a system of two coupled nonlinear ordinary differential equations.  Note that on the left-hand-side we have the change rate of the parameters in terms of $s$, the logarithm of the scale parameter $p$ as defined in (\ref{eq:s-p}).  Thus the right-hand-side can be considered as the two-dimensional "beta function" for our problem, a term used in general for the velocity field of the evolution of coupling parameters in RG methods \cite{ZinnJustin:2005}.

While the mere fact that the ansatz \eqref{eq:FTG-ansatz} is an explicit
solution for RG paths implies that the discontinuity point evolves in $s$ by the
appropriate linear transformation of the coordinate, it is worth checking the
consistency condition \eqref{eq:discontinuity} for it obtained in Sec.\
\ref{S:pde-rg-supp}.  Differentiating \eqref{eq:xextr} by $s$ and substituting
the derivatives from \eqref{eq:eps-gamma} yields
\begin{equation}
 \dot x_{\mathrm{extr}} = - \frac{\epsilon}{\gamma'} \ln \left( 1+
\frac{\gamma'}{\epsilon}\right) -1.  \label{eq:xextr-evolv}
\end{equation}
Using \eqref{eq:xextr} again we recover just the evolution equation
\eqref{eq:discontinuity} of a discontinuity.  Thus the ansatz
\eqref{eq:FTG-ansatz} with a discontinuity at the extremum can be consistently
understood as representing a valid distribution with a discontinuity.

Turning to the solutions of \eqref{eq:eps-gamma}, some of them can be
immediately seen at a glance. Firstly, all fixed points lie on the $\epsilon=0$
fixed line.

Secondly, $\gamma'=0$ is an invariant line, where Eq.\ \eqref{eq:eps}
simplifies to
\begin{equation}
 \dot{\epsilon} = 2\epsilon^2. \label{eq:eps0}
\end{equation}
Note that Eq.\ \eqref{eq:to-ode2} becomes singular for  $\gamma'\to 0$, but if
$\gamma'\equiv 0$ is inserted in \eqref{eq:to-ode1} then we again arrive at
\eqref{eq:eps0}.  Its solution with initial condition $\epsilon_0=\epsilon(0)$
is
\begin{equation}
 \epsilon(s) = \frac{\epsilon_0}{1-2\epsilon_0s}. \label{eq:eps0-solv}
\end{equation}
Thus we obtain the remarkable feature that the FTG fixed point perturbed by the
marginal eigendirection, i.e., \eqref{eq:FTG-ansatz} with $\gamma'=0$, is a
solution not only when the perturbation is small, but also when it is of
arbitrary magnitude.  One should keep in mind, however, that the support of the
distribution is given by \eqref{eq:borders} with
$x_{\mathrm{extr}}=1/\epsilon(s)$, thus in this case it does not extend to the
full real axis.

Thirdly, by substituting  $\epsilon=-\gamma'$ we again arrive at an invariant
line, where
\begin{equation}
 \dot{\epsilon} = \epsilon^2, \label{eq:eps1}
\end{equation}
thus
\begin{equation}
 \epsilon(s) = \frac{\epsilon_0}{1-\epsilon_0s}. \label{eq:eps1-solv}
\end{equation}
This case corresponds to the somewhat pathological  situation discussed in the
end of the previous subsection, where the function  $g(x,s)$ is monotonic over
the entire real axis but does not develop the divergences in both directions
$\pm\infty$ as needed to yield a valid distribution.

In the general case,  we can determine the relation $\epsilon(\gamma')$, whose
differential equation can be obtained from Eq.~\eqref{eq:eps-gamma} as
\be
\frac{d\epsilon}{d\gamma'} =  \frac{2\epsilon}{\gamma'}+ 1.
\ee
The homogeneous part of this linear differential equation has the general
solution $\epsilon_\text{h}= A \gamma'^2$, while a particular solution of the
inhomogeneous equation can be given as $\epsilon_\text{p}= - \gamma'$. Hence in
general
\begin{equation}
 \epsilon = A \gamma'^2 - \gamma', \label{eq:eps-s}
\end{equation}
where the constant is determined  by the initial condition as
\begin{equation}
A = \frac{\epsilon_0+\gamma_0'}{\gamma_0'^2}  \label{eq:eps-s-ic}
\end{equation}
Now we can obtain the differential equation for $\gamma'(s)$ by substituting
\eqref{eq:eps-s}  into \eqref{eq:gammap} as
\begin{equation}
 \dot\gamma' = \gamma'^2 (A \gamma' -1).
\end{equation}
This solves to
\be
s = \frac{1}{\gamma'} - \frac{1}{\gamma'_0} +
\frac{\epsilon_0+\gamma_0'}{\gamma_0'^2}\ln \left(
1+ \frac{\gamma'_0}{\epsilon_0}  -
\frac{\gamma_0'^2}{\epsilon_0\gamma'}
\right), \label{eq:s-ftg}
\ee
where \eqref{eq:eps-s-ic} had been used.  Hence by Eqs.\
(\ref{eq:eps-s},\ref{eq:s-ftg}) we have the sought solution in
the form of explicitly given $\epsilon(\gamma')$ and $s(\gamma')$
functions.

%specifying the relation $\gamma'(s)$, whence by
%(\ref{eq:eps-s},\ref{eq:eps-s-ic}) the $\epsilon(s)$ is determined.

\subsubsection{Trajectories in parameter space}
\label{S:param-ftg1}

From Eq.\ \eqref{eq:eps-s} we see that the RG flow takes place generally on
parabolas in the plane ($\gamma'$,$\epsilon$), with the exception of the
straight lines $\epsilon=-\gamma'$ and $\gamma'=0$.  The first line is obtained
when $A=0$, while the second one for infinite $|A|$ with the limits $A\to \pm
\infty$ corresponding to the positive and negative parts of the ordinate,
respectively.

\begin{center}
\begin{figure}[htb]
\includegraphics[angle=0,width=8.3cm]{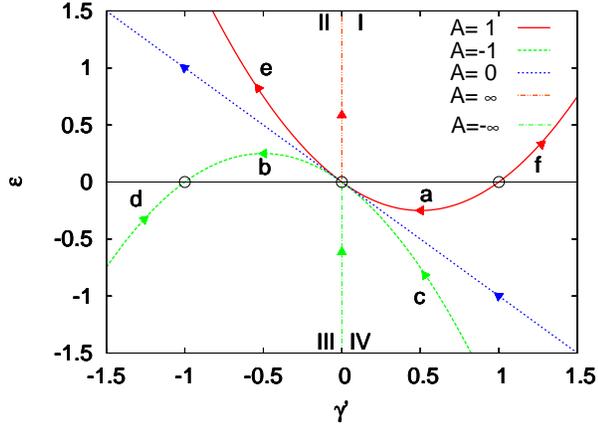}
\caption{Main types of RG paths when the deviation from FTG is given
by a
single eigenfunction as in \eqref{eq:FTF-ansatz}. The characteristic
cases of
$A=-1,0,1$ together with the limits $A=\pm \infty$ are plotted.
Circles mark
their intersections with the fixed line $\epsilon=0$. The segments a,b
represent
closed loops, c,d converge to FTG, whereas e,f diverge from it.
\label{F:RG_flow_ftg}}
\end{figure}\end{center}

Typical paths in parameter space are plotted in Fig.\ \ref{F:RG_flow_ftg}. The
direction of the flow is also marked as determined from \eqref{eq:gammap}: the
$\gamma'$ grows in the quarter planes I,III and decreases in II,IV. The flow
remains in the same quarter-plane as it started from.  The following main cases
can be distinguished (references to the labels of various line segments will be
indicated in parentheses like (a) for segment ``a'' etc.).

\textit{(i) Convergence in the plane -- excursions in function space}: (a) In
quarter IV, with $0<-\epsilon_0<\gamma_0'$, i.e. $A>0$, the distribution
converges to FTG along the marginal direction $\gamma'\to 0$. The initial
condition can be arbitrarily close to FTG, differing from it in an unstable
eigenfunction.  (b) In quarter II, with $0<\epsilon_0<-\gamma_0'$, the path
converges to FTG along a linearly stable direction $\gamma'\to 1/A< 0 $. It can
start arbitrarily close to FTG, differing from it in the marginal eigenfunction.
  In both cases (a,b) Eq.\ \eqref{eq:borders} shows that the support extends to
the real axis.

\textit{(ii) Convergence in the plane -- non-returning solutions  in function
space}:  (c) In quarter IV, with $-\epsilon_0\ge \gamma_0'>0$, i.e.\ $A\leq 0$,
the path converges to FTG along the marginal direction. (d) In quarter III the
trajectory converges to FTG along a linearly stable direction with $\gamma'\to
1/A< 0$.   Equation \eqref{eq:borders} shows that in both cases (c,d), with the
exception of $-\epsilon_0= \gamma_0'$, i.e.\ $A=0$, the support is bounded from
below where the distribution exhibits a jump. This restriction does not affect
the FTG limit for the statistics of the maximum, but the paths diverge in
parameter plane when one follows them backward.  From Eqs.\
(\ref{eq:xextr},\ref{eq:borders},\ref{eq:eps-s}) one sees that along the
divergence the lower border $x_\ast\to 0$ from below.

\textit{(iii) Divergence in the plane -- no convergence to FTG}: (e) In quarter
II, with $\epsilon_0\ge-\gamma_0'>0$ the path diverges.   From
\eqref{eq:borders}  it follows that for $\epsilon_0> -\gamma_0'$, i.e.\ $A>0$,
the support is bounded from above, with the bound $x^\ast$ converging to zero as
it can be seen from Eqs.\ (\ref{eq:xextr},\ref{eq:borders},\ref{eq:eps-s}).
There we understand a jump in the integrated distribution to $1$.   But in the
case when the upper border of a distribution has a finite probabilistic weight,
the limit distribution of the maximum must be  just the discrete distribution at
the limit of the upper border.  This gives now  $x=0$ with probability one.
Since the support is unbounded from below, paths here can be continued backward
along $s\to-\infty$ to FTG, converging to it along the marginal direction. (f)
Starting anywhere in quarter I leads again to a path diverging from FTG,
similarly to case (e) Continuing backward gives FTG, asymptotically along a
linearly unstable direction.

In what follows we will give some explicit examples for the different types of
RG trajectories discussed in this subsection.

\subsubsection{Excursions in distribution space}
\label{S:ustable-ftg1b}

We now illustrate the solution given in the previous subsection on the case of
the returning invariant manifolds discussed in (i) of the previous subsection.
If we start an  excursion near the FTG distribution in a linearly unstable
direction with $1\gg -\epsilon_0 >0$ then the path will turn back to the FTG
fixed point  along the marginal direction.  This corresponds to paths like (a)
on Fig.\ \ref{F:RG_flow_ftg}.   We shall prefer a parameterization of the
trajectory such that the $s\to -\infty$ limit corresponds to
$\gamma'\to\gamma'_{\mathrm{i}}>0$, so by \eqref{eq:eps-s} we have
$\gamma'_{\mathrm{i}}=1/A>0$, it is the initial parameter,  and also
$\epsilon\to\epsilon_{\mathrm{I}}=0$.  For $s\to \infty$ the path goes to the
origin $\epsilon=\gamma'=0$.  There remains  an arbitrariness in choosing the
origin of the parameter $s$: we set it now such that at $s=0$ the $-\epsilon(s)$
attains its maximum. This happens at $\gamma'_0= \gamma'_{\mathrm{i}}/2$ where
$\epsilon_0= -\gamma'_{\mathrm{i}}/4$. Then from \eqref{eq:s-ftg} we get
\be
s = \frac{1}{\gamma'} - \frac{2}{\gamma'_{\mathrm{i}}} +
\frac{1}{\gamma'_{\mathrm{i}}} \ln \left(
\frac{\gamma'_{\mathrm{i}}}{\gamma'}-1\right). \label{eq:s1-ftg}
\ee
From this equation, the asymptotes of $\gamma'(s)$ can be computed as
\renewcommand{\arraystretch}{2}
\begin{equation}
\gamma' \approx  \left\lbrace
\begin{array}{lll}
\gamma'_{\mathrm{i}}(1-e^{-|s|\gamma'_{\mathrm{i}}}), \ &\mathrm{if} \
&s \to
-\infty, \\
\dfrac{1}{s}, \ &\mathrm{if} \  &s \to \infty.\\
\end{array}\right.
\end{equation}

\begin{center}
\begin{figure}[htb]
\includegraphics[angle=0,width=8.3cm]{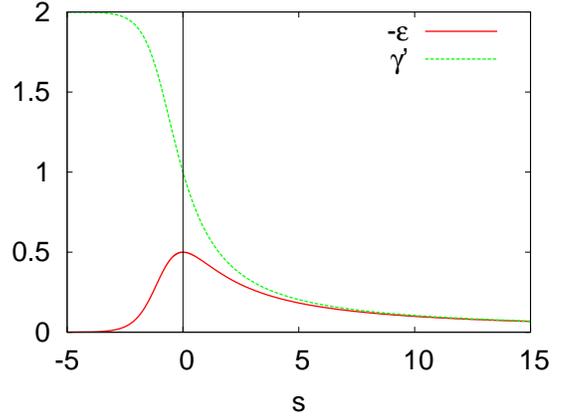}
\caption{Amplitude $\epsilon$ and exponent $\gamma'$  of the unstable
manifold as function of the parameter $s$ for the case
$\gamma'_{\mathrm{i}}=2$.
\label{F:manifolds_pars}}
\end{figure}

\begin{figure}[htb]
\includegraphics[angle=0,width=8.3cm]{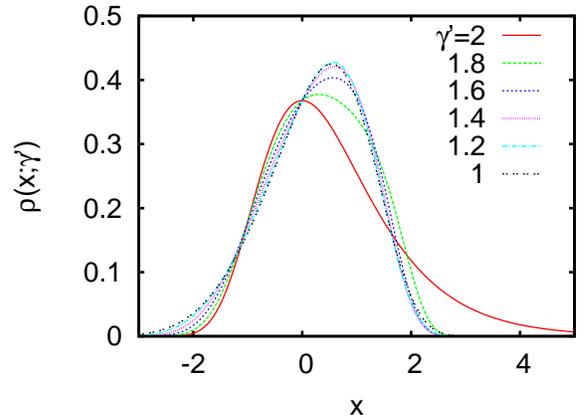}
\caption{Probability densities \eqref{eq:rho-fig} of the unstable
manifold with the parameters given in Fig.\ \ref{F:manifolds_pars}, during $s\le
0$, i.e., $2\ge \gamma'\ge 1$, where the amplitude of the perturbation
$\epsilon$ grows.  The numbers indicated are the $\gamma'$ values, the initial
value $\gamma'_{\mathrm{i}}=2$ corresponds to the FTG distribution, and
$\gamma'=1$ to the farthest deviation from it measured in $\epsilon$.  The
parameters associated with the decreasing sequence of $\gamma'$s are
$s=-\infty, -1.543, -1.0681, -0.7094, -0.36950, 0$
resp. \label{F:manifolds1}}
\end{figure}

 \begin{figure}[htb]
 \includegraphics[angle=0,width=8.3cm]{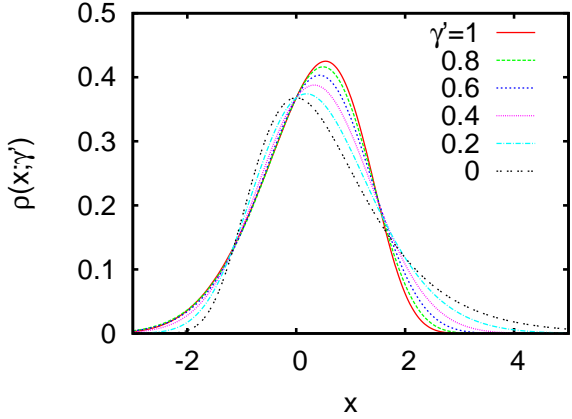}
\caption{Same as in Fig.\ \ref{F:manifolds1}, but during $s\ge 0\,
(0\le \gamma'\le 1)$ where the amplitude of the perturbation  $\epsilon$
decreases. The sequence begins at $\gamma'=1$ and goes to $\gamma'=0$, the
latter corresponding to the FTG distribution because there $\epsilon=0$.  The
parameters associated with the decreasing sequence of $\gamma'$s are $s=0,
0.4527, 1.0903, 2.1931, 5.0986, \infty$, resp.
\label{F:manifolds2}}
 \end{figure}
\end{center}

Figure \ref{F:manifolds_pars} displays the above obtained $\gamma'(s)$ and
$\epsilon(s)$ curves for $\gamma'_{\mathrm{i}}=2$.  We also illustrated the
corresponding one-parameter family of probability densities as defined through
\eqref{eq:FTG-ansatz} by
\begin{equation}
 \rho(x;\gamma') = \partial_x \, e^{-e^{\scriptstyle -g(x,s(\gamma'))}}. \label{eq:rho-fig}
\end{equation}
First we start out from the very proximity of the FTG distribution and move
farther from it, a sequence up to $s=0$ is shown on Fig. \ref{F:manifolds1},
then for positive parameters $s>0$ the $\epsilon$ turns back towards zero and
eventually the FTG distribution is approached, as displayed on Fig.
\ref{F:manifolds2}.

\begin{figure}[htb]
\includegraphics[angle=0,width=8.3cm]{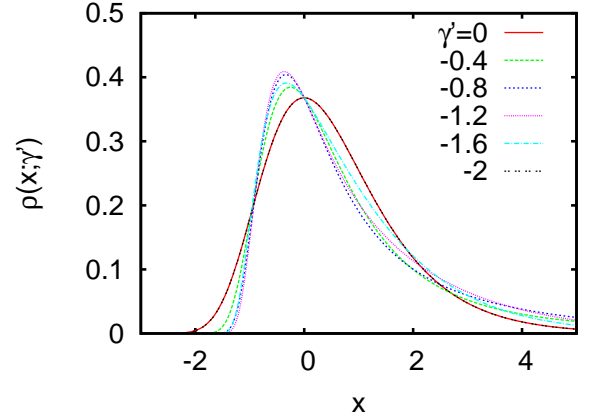}
\caption{The unstable invariant manifold starting near the origin in
parameter plane, corresponding to the path (b) in Fig.\ \ref{F:RG_flow_ftg}.
While $\gamma'$ changes from $0$ to $\gamma_{\mathrm{f}}'=-2$ as marked in the
legend, the distribution diverges from then converges to the FTG fixed point.
\label{F:rgflow-um2}}
 \end{figure}

The other looping invariant manifold corresponds to the line segment (b) in
Fig.\ \ref{F:RG_flow_ftg}. The solution is given by the same formula as
\eqref{eq:s1-ftg} but now with  $\gamma_{\mathrm{i}}'$ replaced by a
$\gamma_{\mathrm{f}}'<0$. While in the parameter plane a simple symmetry
transformation relates segment (a) with (b), the distributions themselves are
manifestly different due to the sign change of the perturbation.   This is
demonstrated on Fig.\ \ref{F:rgflow-um2}, where the functions are oppositely
skewed than those in the excursion on Figs.\
\ref{F:manifolds1},\ref{F:manifolds2}.

\subsubsection{Non-returning paths in distribution space}
\label{S:nonreturn-ftg1}

We shall illustrate paths not making excursions from and to FTG, as described in
paragraphs (ii) and (iii) of subsection \ref{S:param-ftg1}. The solution
(\ref{eq:eps-s},\ref{eq:s-ftg}) is valid, with  semi-infinite support for
$\epsilon\neq 0$, whence the continuous part $\rho_{\mathrm c}(x;\gamma')$ of
the probability density function can be reconstructed.
\begin{center}
\begin{figure}[htb]
\includegraphics[angle=270,width=8.3cm]{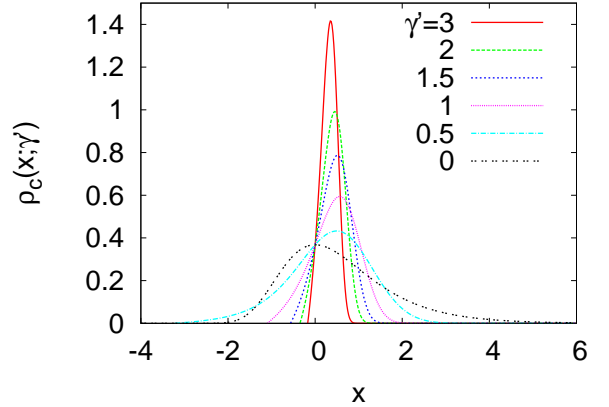}
\caption{Continuous part of the density  for the trajectory starting
in quarter IV in parameter plane and converging to FTG, as it corresponds to the
path (c) in Fig.\ \ref{F:RG_flow_ftg}. The integral of this function is less
than one, except in the FTG limit $\gamma'=0$. Characteristic values of
$\gamma'$ of the distributions are displayed, converging to zero while
$s\to\infty$.
\label{F:rgflow-conv}}
 \end{figure}

\begin{figure}[htb]
\includegraphics[angle=270,width=8.3cm]{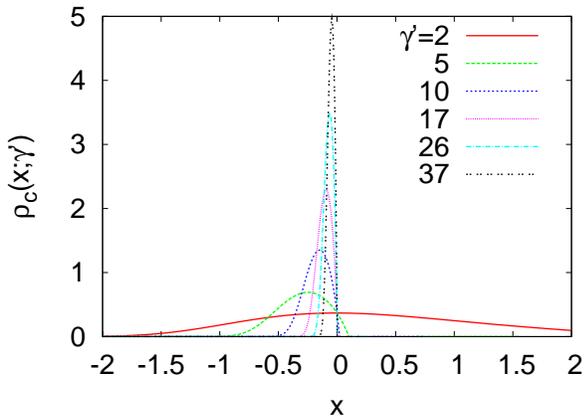}
\caption{Continuous part of the density  for the trajectory starting
in quarter I  and diverging in the parameter plane, as it corresponds to the
path (f) in Fig.\ \ref{F:RG_flow_ftg}. The integral of this function is less
than one, except in the FTG limit, obtained by backtracking for $s\to-\infty$.
Characteristic values of $\gamma'$ of the distributions are displayed, where
$\gamma'=2$ corresponds to FTG, as seen from  Fig.\
\ref{F:RG_flow_ftg}.
\label{F:rgflow-div}}
 \end{figure}
\end{center}

First we consider a path corresponding to the segment (c) of Fig.\
\ref{F:RG_flow_ftg}, where \eqref{eq:borders} gives a finite lower border
$x_\ast$  for $\gamma'>0$.  Then the continuous part of the density
$\rho_{\mathrm c}(x;\gamma')$ vanishes at $x_\ast$ and has a norm smaller than
one, as illustrated by the sequence in Fig.\ \ref{F:rgflow-conv}.   As explained
in Sec.\ \ref{S:ansaty-ftg1} we associate a complementary discrete probabilistic
weight with $x_\ast$.  This weight, never exceeding $1/e$ and vanishing in the
FTG limit, could be represented by a Dirac delta in the density, but we omit it
from the figure. Secondly, we illustrate in Fig.\ \ref{F:rgflow-div} a
trajectory of the segment (f) of Fig.\ \ref{F:RG_flow_ftg}, where
\eqref{eq:borders} gives a finite upper border $x^\ast$.  For $s\to-\infty$ we
have the FTG limit, and for increasing $s$ the $x^\ast$ decreases and reaches
zero. This happens at a finite $s$, as it can be seen from (\ref{eq:s-ftg}),
corresponding to the point at infinity along the parabola segment (f) on Fig.\
\ref{F:RG_flow_ftg}.  In this limit the non-normalized part of the distribution
becomes peaked at the origin, and, together with the discrete weight at the
upper border of the support (not shown on Fig.\ \ref{F:rgflow-div}), form a
Dirac delta at the origin.  This single weight is the generic EVS limit
distribution when the parent has a discrete weight at the upper border of the
support, and formally corresponds to the $\gamma\to -\infty$ limit of the FTW
fixed point distribution.

The above illustrated cases are the typical non-returning trajectories,
converging ones go to FTG, while diverging ones in the plane represent
convergence to a Dirac delta.

This ends our examples where a single eigenfunction, appropriately
parameterized, gives exactly the RG path even farther away from the FTG fixed
point.

%%%%%%%%%%%%%%%%%%%%%%%%%%%%%%%%%%%%%%%%%%%%%%%%%%%%%%%%%%%%%%%%%%%%%%%%
\subsection{RG flow in terms of the general linearized solution around FTG}
\label{S:ustable-ftg2}
%%%%%%%%%%%%%%%%%%%%%%%%%%%%%%%%%%%%%%%%%%%%%%%%%%%%%%%%%%%%%%%%%%%%%%%%

\subsubsection{Reparameterization of the general solution of the linearized PDE}
\label{S:ustable-ftg2-1}

Given the RG trajectories in the previous subsection \ref{S:ustable-ftg}, whose
functional form is characterized by a single eigenfunction, it is natural to ask
whether the solution could be generalized to linear combinations of the
eigenfunctions. In other words, assuming the form of the general linearized
solution as described in Sec.\ \ref{S:gen-lin-sol}, for the FTG case, the
question is whether we can parameterize it appropriately such that it provides
exact RG paths.

Considering the explicit form of the FTG eigenfunction \eqref{eq:psi-FTG}, one
sees that $\gamma'$ acts as a rescaling of the variable $x$, as well as a global
rescaling factor of the function $\psi$. Noticing that the variations of this
global rescaling factor can be reabsorbed into $\epsilon$ through a suitable
redefinition of $s$, we conclude that the ansatz \eqref{eq:FTG-ansatz} is
equivalent to a rescaling of $x$ together with a reparameterization of $s$. This
suggests to use the following more general ansatz, where the eigenfunction is
now replaced by a general solution $\phi$ of the linearized equation as
\be \label{eq:sol-gen-nonlin}
g(x,s) = x + \phi\big(\beta(s)x,\eta(s)\big),
\ee
where $\beta(s)$ and $\eta(s)$ are functions to be determined, and
$\phi$ is defined in (\ref{sol-gen-eqlin}) for $\gamma=0$ as
\begin{equation}
 \phi(x,s) = \chi(s) + x \chi'(s) - \chi(x+s).\label{eq:phi-ftg}
\end{equation}
Inserting the above form of $g(x,s)$ into the PDE (\ref{eq:pde}) yields a set of
two coupled differential equations for $\beta(s)$ and $\eta(s)$:
\begin{subequations}\label{eq:eta-beta}
\begin{align}
\dot{\eta} &= \beta, \label{eq:eta-dot} \\
\dot{\beta} &= \beta^3 \chi''(\eta). \label{eq:beta-dot}
\end{align}
\end{subequations}
Consequently, the ansatz (\ref{eq:sol-gen-nonlin},\ref{eq:phi-ftg}) is
an exact RG trajectory if the running parameters satisfy the above ODE's.
This is remarkable if one recalls that it represents an extension of the
solution in the linear neighborhood of the FTG fixed point by only using a
suitable parameterization.  Even more remarkably, as we shall see it below,
the present ansatz is actually the most general solution of the PDE for the RG
flow, and the meaning of the parameter functions will also be revealed.

Before going on, we should note that the pair of ODE's \eqref{eq:eta-beta} can
be solved explicitly for the $\beta(\eta)$ and $s(\eta)$ functions with
arbitratry initial conditions. We can obtain this solution by firstly dividing
the two equations to yield an ODE for $\beta(\eta)$, which can be integrated,
and secondly, by substituting the thus obtained $\beta(\eta)$ into
\eqref{eq:eta-dot}, which then yields explicitly $s(\eta)$.  Instead of
displaying and analyzing this solution, however, we shall follow a different and
perhaps more surprising reasoning.

\subsubsection{ODE's for a general RG path}
\label{S:ustable-ftg2-2}

Let us reconsider the  scale and shift parameters of the RG transformation,
i.e., $a(s)$ and $b(s)$ of Sec.\ \ref{S:pde-rg}.  From their definition
Eqs.\ (\ref{eq:bs},\ref{eq:as}) we obtain a pair of ODE's
\begin{subequations} \label{eq:a-b-ode}
\begin{align}
 \dot b &= a, \label{eq:b-dot}\\
 \dot a &= -a^3 \, g''(b), \label{eq:a-dot}
\end{align}
\end{subequations}
where we remind the reader that $g(x)=g(x,0)$ is the initial condition for the
PDE, corresponds to the parent distribution, and meets the standardization
conditions \eqref{eq:std-g}.  The above equations are essentially the same as
the formerly displayed (\ref{eq:a-bdot},\ref{eq:gamma-s}).  The initial
condition of the scale and shift parameters is obviously $a(0)=1$ and $b(0)=0$.

Conversely, it is straightforward to show that the above pair of ODE's, with the
aforementioned initial condition, can be uniquely solved resulting in
(\ref{eq:bs},\ref{eq:as}).

We can immediately recognize the similarity between the pairs of equations
\eqref{eq:eta-beta} and \eqref{eq:a-b-ode}.  There are still a few steps
to make before the analogy is fully specified.  First, we should set the initial
conditions in \eqref{eq:eta-beta} to $\eta(0)=0$ and $\beta(0)=1$.  Then, using
these values, we obtain the initial function $g(x)$ in terms of $\chi(x)$ by
substitution of  \eqref{eq:phi-ftg}  into \eqref{eq:sol-gen-nonlin} to yield
\begin{equation}
 g(x) = g(x,0)= x + \chi(0) + x \chi'(0) - \chi(x).  \label{eq:g-chi}
\end{equation}
Hence  $g''(x)=-\chi''(x)$ and we see that the two sets of ODE's
\eqref{eq:eta-beta} and \eqref{eq:a-b-ode} are identical.  Therefore,
the correspondence
\begin{subequations}
\begin{align}
\eta(s)&=b(s)=g^{-1}(s)  \label{eq:eta-b}\\
\beta(s)&=a(s)=1/g'\big(\eta(s)\big) \label{eq:beta-a}
\end{align} \label{eq:eta-b-beta-a}
\end{subequations}
holds.  Eventually, it remains to be clarified what
the role of the $\eta$ and $\beta$ parameters in formula
\eqref{eq:sol-gen-nonlin} for the RG flow is. While the parameters $a(s)$ and
$b(s)$ enter as the respective scale and shift in the argument of
\eqref{eq:gxs}, at first glance it is not obvious that also $\eta$ and $\beta$
in \eqref{eq:sol-gen-nonlin} define a linear transformation. We start out from
\eqref{eq:gxs}, substituting the  function $g(x)$ from \eqref{eq:g-chi} and
$s(\eta)$ from the inversion of \eqref{eq:eta-b}.  This yields
\begin{align}
g(x,s)=& g\big(\beta(s)x+\eta(s)\big)-s \nonumber \\
=& \beta x +\eta + \chi(0) + ( \beta x +\eta ) \chi'(0)- \chi(\beta x +\eta)
\nonumber \\&- \eta - \eta\chi'(0) -\chi(0) + \chi(\eta)  \nonumber \\
=& \beta x \big(1+ \chi'(0)\big) + \chi(\eta) - \chi(\beta x + \eta).
\label{eq:equivalence1}
 \end{align}
Equations \eqref{eq:g-chi} and \eqref{eq:beta-a} have the consequence
\be
\beta \big(1+ \chi'(0)\big) = 1 + \beta \chi'(\eta),
\ee
which can be substituted into \eqref{eq:equivalence1} to yield
\be
\begin{split}
g(x,s)=& x  +  \beta x \chi'(\eta) + \chi(\eta) - \chi(\beta x +
\eta)\\
=& x + \phi(\beta x,\eta).
\end{split}
\label{eq:equivalence2}
\ee
The last line is obtained by the definition \eqref{eq:phi-ftg}, whereby the
starting ansatz  \eqref{eq:sol-gen-nonlin} is recovered.

In summary, the trial function \eqref{eq:sol-gen-nonlin} includes the most
general, defining formula \eqref{eq:gxs} for the RG trajectories.   A
direct derivation of this conclusion, without using the differential formalism,
is presented in Appendix \ref{S:ustable-ftg2-4}.   One should recall
that the function $\chi(x)$ in \eqref{eq:phi-ftg}, if small,
is related to the weight in a linear combination of eigenfunctions, as seen in
Sec.\ \ref{S:pert-fp-pde}.  In fact, the motivation for \eqref{eq:sol-gen-nonlin} was to construct a generalization  of the solution in the linear neighborhood of the FTG fixed point, valid also in the non-perturbative regime, i.e., for not only small $\chi(x)$.  While the RG paths in terms of a single eigenfunction about FTG, as discussed in Sec.\ \ref{S:ustable-ftg}, could be considered as a curiosity, now we see that the RG transformation generally maintains the functional form also for linearly weighted combination of eigenfunctions, if appropriately parameterized, even farther from the FTG fixed point.  This is to our knowledge a very special property for an RG transformation, unparalleled over its areas of applications in statistical physics.

\subsubsection{ODE's for the perturbation parameter}
\label{S:ustable-ftg2-3}

In order to make a connection with the single eigenfunction case, as well as
to explicitly characterize the deviation from the FTG fixed point, a
perturbation amplitude $\epsilon$ can be introduced for a general RG trajectory.
With the notation of Sec.\ \ref{S:ustable-ftg2-1} let us define
\begin{equation}
 \epsilon = \beta^2 \chi''(\eta).\label{eq:eps-b}
\end{equation}
Hence we can rewrite \eqref{eq:beta-dot} as
\begin{equation}
  \dot\beta = \epsilon\, \beta. \label{eq:a-dot2}
\end{equation}
Taking the derivative of \eqref{eq:eps-b} with respect to $s$, we get
using the pair of ODE's \eqref{eq:eta-beta}
\be
 \dot \epsilon = 2 \epsilon^2 + \beta^3 \, \chi'''(\eta).
\ee
Then reexpressing $\eta$ from \eqref{eq:eps-b}, we can write $\dot\epsilon$
as a function of $\epsilon$ and $\beta$ only
\begin{equation}
 \dot \epsilon = 2 \epsilon^2 + \beta\, \epsilon\,
\varphi(\epsilon/\beta^2),\label{eq:eps2}
\end{equation}
where we have introduced the auxiliary function
\be
\varphi(y) = \frac{\chi'''(\chi''^{-1}(y))}{y}.
\ee
The ODE's  (\ref{eq:a-dot2},\ref{eq:eps2}) can be considered as the
generalization of \eqref{eq:eps-gammap-ode}, valid for RG paths in terms of a
single eigenfunction.   The latter is recovered as a special case in the
following way.  Consider the initial conditions $\beta(0)=1$,
$\gamma'(0)=\gamma_0'$, and $\epsilon(0)=\epsilon_0$, then  identify $\gamma'$
with $\beta\gamma_0'$,  choose $\chi(x)$ so that $\chi''(x)=\epsilon_0
e^{\gamma_0'x}$ yielding $\varphi(y)\equiv \gamma_0'$,  thus we are led back to
\eqref{eq:eps-gammap-ode}.

Note that the parameter $\epsilon(s)$ introduced above equals the ``effective''
$\gamma(s)$ of \eqref{eq:gamma-s}, and can be considered as a measure of the
deviation from the FTG fixed point in the general case, not only when a
deviation is represented by a single eigenfunction.

\subsubsection{An example}
\label{S:ustable-ftg2-3b}

We shall illustrate the RG paths as given in the previous part of
Sec.\ \ref{S:ustable-ftg2} on an example for an excursion from the FTG fixed
point.   Note that returning invariant manifolds were, remarkably,
given in closed forms in Sec.\ \ref{S:ustable-ftg}.  The example shown here can
be considered as a generalization of them.   Consider a linear combination of
unstable eigenfunctions, weighted uniformly between $\gamma'=0$
and a $\gamma'_0>0$, corresponding to
\begin{equation}
 \chi(x) = \frac{3\epsilon_0}{\gamma_0'^3}\int_0^{\gamma_0'}
e^{\gamma' x}
d\gamma' =  \frac{3\epsilon_0}{\gamma_0'^3}
\frac{e^{\gamma_0'x}-1}{x}.\label{eq:chi2}
\end{equation}
Note that the constant  $\gamma'_0$ now is not to be confounded with an
initial condition, rather it is the upper border of set of $\gamma'$ indices
here.

\begin{center}
\begin{figure}[htb]
\includegraphics[angle=0,width=8.3cm]{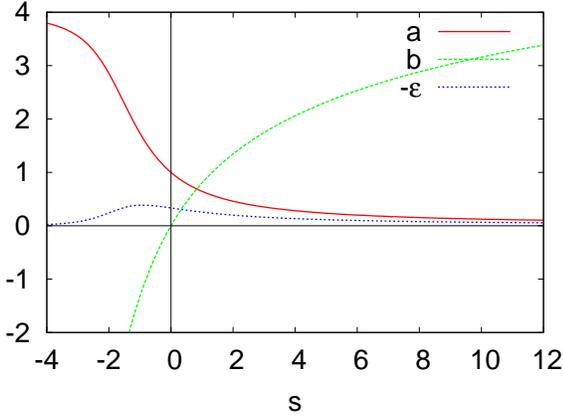}
\caption{Excursion from FTG starting from a linear combination of
eigenfunctions. In \eqref{eq:g-chi} the function \eqref{eq:chi2} with
$\epsilon_0=-1/2$ and $\gamma_0'=1$ was taken.  The running parameters
$a(s)=\beta(s)$ and $b(s)=\eta(s)$ in \eqref{eq:eta-b-beta-a}, and  the
effective $\epsilon(s)$ defined in \eqref{eq:eps2} are displayed. Note that
while Fig.\ \ref{F:manifolds_pars} characterizes single invariant manifolds,
these curves parameterize a continuum bundle of manifolds. The corresponding
probability density function sequence first diverges from then converges to the
FTG fixed point, in a  manner qualitatively similar to the sequences displayed
in Figs.\
\ref{F:manifolds1},\ref{F:manifolds2}.
\label{F:gen_sol_pars}}
\end{figure}
\end{center}

Choosing $\chi(x)$ as given in \eqref{eq:chi2} yields through
\eqref{eq:g-chi} a parent $g(x)$ with support extending to the real axis, if
$2\gamma_0'>-3\epsilon_0>0$.  The scale and shift parameters $\beta=a$, $\eta=b$
are determined through \eqref{eq:eta-b-beta-a} and the perturbation
parameter $\epsilon$ through \eqref{eq:eps-b}.  This is illustrated on Fig.\
\ref{F:gen_sol_pars} whose qualitative similarity with Fig.\
\ref{F:manifolds_pars} is apparent. While this example is about a case when the
support is the real axis, the general formulas in this section also describe
cases with limited support, similarly to the situation when the deviation from
the FTG fixed point was given in terms of a single eigenfunction, as discussed
in Sec.\ \ref{S:ustable-ftg}.

%%%%%%%%%%%%%%%%%%%%%%%%%%%%%%%%%%%%%%%%%%%%%%%%%%%%%%%%%%%%%%%%%%%%%%%%
\subsection{Paths about the FTF and FTW fixed points in terms of
single eigenfunctions}
\label{S:ustable-ftf}
%%%%%%%%%%%%%%%%%%%%%%%%%%%%%%%%%%%%%%%%%%%%%%%%%%%%%%%%%%%%%%%%%%%%%%%%

\subsubsection{Trial functions for the RG path}

It is a natural question to ask whether a generalization of the ansatz
\eqref{eq:FTG-ansatz} in terms a single eigenfunction may give trajectories
belonging to fixed points other than the FTG one.
We propose such a generalization in the form of the trial function
\be
g(x,s) = f\Big( x + \epsilon(s)
\psi\big(x;\bar\gamma(s),\gamma'(s)\big);\bar{\bar\gamma}(s) \Big),
\label{eq:FTF-ansatz}
\ee
where the fixed point function $f$ is given by \eqref{eq:pde-fp}, the
eigenfunction $\psi$ by \eqref{eq:psi-rg}, and the  running parameters
$\epsilon(s),\bar\gamma(s),\gamma'(s),\bar{\bar\gamma}(s)$ are to be determined
from the PDE \eqref{eq:pde}.  We used here the notation $\bar\gamma(s)$ and
$\bar{\bar\gamma}(s)$ to distinguish them from the ``effective'' $\gamma(s)$
given in \eqref{eq:gammas}.  While the  $\bar\gamma$  should be equal to
$\bar{\bar\gamma}$ for a proper eigenfunction about the fixed point with
$\bar{\bar\gamma}$, we allow them to be different for the present trial
function.

After substitution into the PDE \eqref{eq:pde} a somewhat tedious but
straightforward calculation shows that such a proposition can be valid
only if
\begin{subequations}
\begin{align}
\bar{\bar\gamma}(s) &= \gamma_0 ,\\
\gamma'(s)&= B\,\bar\gamma(s), \label{eq:B}
\end{align}
\end{subequations}
where $ \gamma_0$ and $B$ are constants.  Thus the ansatz reduces to
\be
g(x,s) = f\Big( x + \epsilon(s)
\psi\big(x;\bar\gamma(s),B\bar\gamma(s)\big);\gamma_0 \Big),
\label{eq:FTF-ansatz2}
\ee
wherein two running parameters remained. These, again by \eqref{eq:pde}, can be
shown to lead to an RG path if they satisfy the ODE's
\begin{subequations}
\label{eq:eps-gam}
\begin{align}
\dot \epsilon &= 2\epsilon^2 + \epsilon(\gamma_0+(B-1)\bar\gamma),\\
\dot{\bar\gamma} &= \bar\gamma(\epsilon + \gamma_0- \bar\gamma).
\end{align}
\end{subequations}
Consistency with the flow about the FTG fixed point can be checked, if we take
$B\to \infty$ while keeping $\gamma'(s)$ in \eqref{eq:B} finite, and set
$\gamma_0=0$, when we immediately recover the pair of ODE's in
\eqref{eq:eps-gamma}.

An interesting feature emerges if we keep $\gamma_0$ nonzero while $B\to \infty$.  Then starting near the non-FTG fixed point $f(x;\gamma_0)$ but using an eigenfunction belonging to FTG, $\psi(x;0,\gamma'(s))$, also leads to an exact, nonperturbative RG path in function space.

Another remarkable property of the ansatz \eqref{eq:FTF-ansatz2} is that it can also describe fixed points corresponding to $\gamma \neq \gamma_0$.  Indeed, it is straightforward to show that
\be
f(x; \gamma+\epsilon) = f(x+\epsilon \psi(x; \gamma+\epsilon,-\epsilon); \gamma), \label{eq:fp-ef}
\ee
which can be considered as a special structural relation between the fixed point functions and the eigenfunctions.

\subsubsection{Fixed points and flow lines in parameter plane}

In what follows we shall briefly review some basic properties of the above
ODE's.  There are in general four fixed points for nonzero $\gamma_0$.

(i) $\epsilon^\ast= \bar\gamma^\ast =0$, corresponding to the fixed point
function $f(x;\gamma_0)$.

(ii) $\epsilon^\ast=0$, $\bar\gamma^\ast =\gamma_0$,
again with $f(x;\gamma_0)$.

(iii) $\epsilon^\ast=-\gamma_0/2$, $\bar\gamma^\ast
=0$, with $f(x;\gamma_0/2)$.

(iv)  $\bar\gamma^\ast=\gamma_0/(B+1)$,
$\epsilon^\ast=-B\bar\gamma^\ast$, with $f(x;\gamma_0/(B+1))$.
Note that in (iii-iv) the fixed point functions $f(x;\gamma)$ with
$\gamma \neq \gamma_0$ arise as made possible by the special property
\eqref{eq:fp-ef}.

For general flow lines in the parameter plane $\bar\gamma,\epsilon$ an ODE is
obtained from \eqref{eq:eps-gam} as
\begin{equation}
 \epsilon'( \bar\gamma)= \frac{2\epsilon}{\bar\gamma} +
\frac{\epsilon}{\bar\gamma}
\frac{\bar\gamma(B+1)-\gamma_0}{\epsilon+\gamma_0-
\bar\gamma}.
\end{equation}
Some invariant lines can be explicitly found as follows.

(a) $\epsilon\equiv 0$ corresponding to staying at $f(x;\gamma_0)$.

(b) $\bar\gamma\equiv 0$ implying a flow by $\dot \epsilon = 2\epsilon^2
+ \epsilon\gamma_0$,  and containing the fixed points (i,iii).

(c) $\epsilon=-B\bar\gamma$, along which we have $\dot{\bar\gamma} =
\bar\gamma( \gamma_0- (B+1)\bar\gamma)$, containing the fixed points (i,iv).

(d) $\epsilon=(B+1) \bar\gamma(\bar\gamma/\gamma_0-1)$, containing the
fixed points (i,iii,iv).

\begin{center}
\begin{figure}[htb]
\includegraphics[angle=0,width=8.3cm]{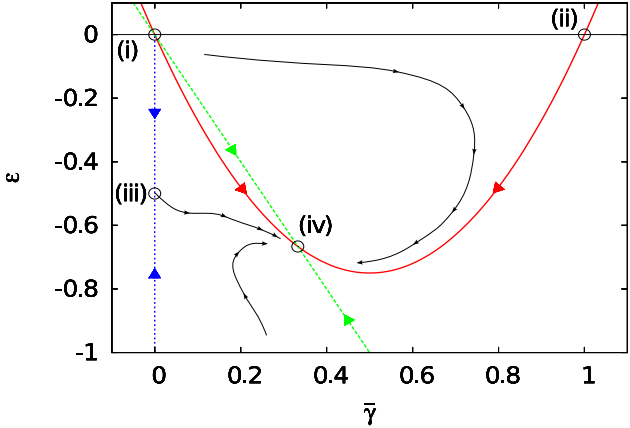}
\caption{RG flow in $\bar \gamma,\epsilon$ space, when the deviation
from the FTF fixed point with  $\gamma_0=1$ is given by a single
eigenfunction as in \eqref{eq:FTF-ansatz2}, with  $B=2$. Circles mark
the fixed points (i-iv) listed in the text, and invariant curves of
the simple types (b-d) as well as some qualitatively drawn flow lines
are indicated. } \label{F:RG_flow_ftf}
\end{figure}\end{center}

For graphical illustration let us restrict ourselves to positive $\gamma_0$ and
$B$.  Then all fixed points lie in the fourth quarter of the parameter plane
with $\bar\gamma\ge 0$, $\epsilon\le 0$,  the (i) is unstable, (ii,iii) are
saddle points, and (iv) is stable.  The latter attracts all RG trajectories
starting within the quarter plane with $\bar\gamma> 0$, $\epsilon< 0$.  On Fig.\
\ref{F:RG_flow_ftf}, we display some characteristic flow lines. The main
difference with respect to the RG paths about the FTG fixed point is that in
this example we do not have excursions, rather, the RG transformation connects
different fixed points.

\subsubsection{Examples}

In order to highlight the above observations we give two families of
solutions.  On the one hand, the family
\be
g(x,s)= f\left( x+\frac{\epsilon}{\gamma'^2} \left(
1+\gamma'x-e^{\gamma'x}\right);\gamma_0\right)
\ee
corresponds to $B\to\infty$ while $B\bar\gamma=\gamma'$ and so from
\eqref{eq:eps-gam} the ODE' become
\begin{subequations} \label{eq:eps-gam2}
\begin{align}
\dot \epsilon &= 2\epsilon^2 + \epsilon(\gamma_0+\gamma'),\\
\dot{\gamma'} &= \gamma'(\epsilon + \gamma_0).
\end{align}
\end{subequations}
Note that this is the generalization of \eqref{eq:eps-gammap-ode} to nonzero
$\gamma_0$.  As the second example we display the RG path along the invariant
line $\epsilon=-B\bar\gamma$ for which,  after some rearrangements, we get
\be
g(x,s)= f\left( \frac{(1+\bar\gamma
x)^{B+1}-1}{\bar\gamma(1+B)};\gamma_0\right).
\ee
The running parameter $\bar\gamma(s)$ satisfies the ODE given under paragraph
(c).  These two examples demonstrate manifestly that eigenfunctions of some
fixed point composed with another fixed point function can lead to exact
trajectories.

\subsubsection{Paradox: can a fixed point with an eigenfunction of another
fixed point be a solution?}

We end this part by calling the reader's attention to a seemingly paradoxical
situation, namely, that \eqref{eq:FTF-ansatz2} contains a fixed point function
with parameter $\gamma_0$, while the eigenfunction within can have, and
typically has, a different $\bar\gamma\neq\gamma_0$ parameter.  Then the
question arises, how a solution near a fixed point, i.e. having a small
$\epsilon$ can be expressed in terms of a non-proper eigenfunction.  The
resolution of that is simple:  If the eigenfunction contains parameters
changing in $s$ then it may be of the functional form of the eigenfunction, but
the perturbation in the argument corresponding to \eqref{eq:g-psi1} does not
separate as \eqref{eq:g-psi} into an $s$- and $x$-depending function. So the
eigenfunction with changing parameters is not a solution of the eigenvalue
problem, thus there is no contradiction with the known solutions.  Only if
the $s$-dependence of the parameter $\bar\gamma$ in \eqref{eq:FTF-ansatz2} is
negligible (vanishing with $\epsilon$) will we obtain separability as in
\eqref{eq:g-psi}.  The parameter  $\bar\gamma$ can change slowly only near
fixed points in the parameter plane, among which we are interested now in
(i) and (ii) located on the $\epsilon=0$ axis.  But then near the fixed point
(i) the eigenfunction becomes
$\psi\big(x;\bar\gamma(s),B\bar\gamma(s)\big)\approx -x^2/2$, a valid
eigenfunction of $f(x;\gamma_0)$ of index $\gamma'=\gamma_0$, furthermore, near
the fixed point (ii) the parameter $\bar\gamma(s)\approx \gamma_0$ so again we
are facing a proper eigenfunction (with $\gamma'=B\gamma_0$).  In summary, the
exact solution \eqref{eq:FTF-ansatz2} does not include a case where an improper
eigenperturbation would appear.

%%%%%%%%%%%%%%%%%%%%%%%%%%%%%%%%%%%%%%%%%%%%%%%%%%%%%%%%%%%%%%%%%%%%%%
%%%%%%
\section{RG for central limit distributions}
\label{S:clt}
%%%%%%%%%%%%%%%%%%%%%%%%%%%%%%%%%%%%%%%%%%%%%%%%%%%%%%%%%%%%%%%%%%%%%%
%%%%%%

In this section we demonstrate a remarkable albeit simple analogy between
limiting behavior in EVS and central limit distributions.  So far we have studied the RG transformation that arose from the EVS problem, as discussed in
Secs.\ \ref{S:basic} and \ref{S:rgflow}.  In its original form
\eqref{eq:def-rg-former} the transformation consists of a linear change of
variable and raising the distribution function to a power.  It is well known,
however, that formally a similar operation, rescaling and raising to power,
should be performed also in the context of the central limit problem on the
moment generating function \cite{Kolmogorov:1954,Christoph:1993}.  In this
section we briefly review this similarity and show that some of the so far considered RG
fixed points actually correspond to distributions known as ``stable'', i.e.,
having the property that sums of random variables keep the same distribution up
to a scale transformation.  The main idea of the RG for distributions falling
into the Gaussian class has been expounded in \cite{JonaLasinio:2001}, treated
as an introductory example, where the eigenvalue problem for analytic
characteristic functions has also been determined.

Below we show that the RG flow equation introduced in this paper for EVS leads,
with a slightly different standardization, to a PDE whose fixed point
solution corresponds to central limit distributions.  We restrict ourselves for
the sake of simplicity here to real, symmetric moment generating functions, so
the fixed point will describe probability densities with even symmetry.
We also evaluate the eigenfunctions and again find a close similarity to those
found in EVS.

\subsection{RG transformation}

Consider the random variable $Z$ defined as a rescaled
sum of random variables, namely
\be
 Z = \frac{1}{a_N}\sum_{i=1}^{N} z_i,
\ee
where $z_i$ are i.i.d.\ numbers each with density $P(z)$ and moment
generating function
\be
\Phi(q) = \int_{-\infty}^{\infty} \td z \, e^{iqz}\, P(z),
\label{eq:mom-gen}
\ee
and $a_N$ is a suitable scaling factor ensuring a non-degenerate limit
distribution, if such exists.
We assume that $P(z)=P(-z)$, which implies $\langle z \rangle=0$.
From the parity of $P(z)$, we also have that the moment generating
function $\Phi(q)$ is real, and satisfies $\Phi(q)=\Phi(-q)$.
Then the moment generating function for the variable $Z$ is known to
be
\be
\Phi_N(q) = \Phi^N(a_Nq),
\ee
also real and symmetric.  Allowing a continuation of $N$ and
introducing $s=\ln N$ we suitably define a RG transformation on the moment
generating function as
\be
\Phi(q,s) = [\hat{R}_s \Phi]  (q)
= \Phi^{e^s}\! \big(a(s)\,q\big).
\label{eq:def-rg-clt}
\ee
An alternative interpretation of the RG would also be to perform a decimation operation,
similarly to the procedure presented in Sec.~\ref{S:basic} for EVS.
The function $\Phi(q,s)$ is real and even in $q$, and is related to the
corresponding probability
density by a Fourier transformation like in \eqref{eq:mom-gen}.
While $\Phi(0,s)\equiv 1$
follows from \eqref{eq:mom-gen}, we have the freedom to impose a
standardization condition, which sets the scale factor $a(s)$, so we require
\be
\Phi(1,s)=\Phi(-1,s)\equiv e^{-1}. \label{eq:stand-clt}
\ee
The above transformation is in clear analogy to the RG operation for EVS
\eqref{eq:def-rg}, but while RG acted on the integrated distribution function
there, here its argument is the moment generating function. Note the difference
in standardization:  in EVS the distribution function became $1$ at the upper
limit of the support and was $e^{-1}$ at $x=0$, now $1$ is reached in the origin
and $e^{-1}$ at $q=-1$. Another difference is that we do not have a monotonicity
condition in $q$ here, nonetheless, the analogy between \eqref{eq:def-rg} and
\eqref{eq:def-rg-clt} is obvious.

To further emphasize the similarity to the RG equation in EVS, we
introduce the function $h(q,s)$ by
\be
\Phi(q,s) = e^{-e^{\scriptstyle -h(q,s)}}.\label{eq:def-h-clt}
\ee
Then $h(q)=h(q,0)$ corresponds to the starting moment generating
function $\Phi(q)=\Phi(q,0)$. Since $h(q,s)$ is even in $q$, it suffices to
only consider one semi-axis in $q$.  To prepare for the analogy to
EVS, we shall consider the negative semi-axis, i.e.\ $q<0$.  The RG
transformation \eqref{eq:def-rg-clt} implies
\be
h(q,s) = h\big(a(s)\,q\big) -s \label{eq:h-q-s}.
\ee
From $\Phi(0,s)\equiv 1$ and the standardization \eqref{eq:stand-clt}
follow
\begin{subequations}
\begin{align}
h(0,s) &\equiv \infty, \\
h(-1,s)&\equiv 0,
\end{align}
\end{subequations}
the latter setting the scale parameter by \eqref{eq:h-q-s} as
\be
h(-a(s))=1+s.
\ee

Differentiation of \eqref{eq:h-q-s} by both variables yields
\begin{subequations}
\begin{align}
\partial_sh(q,s) &=q\, \dot a(s)\, h'(a(s)q)-1, \\
\partial_qh(q,s)&=a(s)\, h'(a(s)q).
\end{align}
\end{subequations}
Hence we obtain the PDE for $h$ as
\be
\partial_sh(q,s) = q\, \gamma(s)\, \partial_qh(q,s)
-1,\label{eq:pde-clt}
\ee
where
\be
\gamma(s)=\frac{\dot a(s)}{a(s)}.
\ee
The standardization condition \eqref{eq:stand-clt} implies that the
left-hand-side of the \eqref{eq:pde-clt} is zero at $q=- 1$, whence
\be
\gamma(s)=-\frac{1}{\partial_qh(-1,s)}.\label{eq:gamma-clt}
\ee
Thus we wound up in \eqref{eq:pde-clt} with a nonlinear PDE, which
resembles the PDE \eqref{eq:pde} obtained for the EVS problem, while the
analog of \eqref{eq:gamma-clt} is \eqref{eq:gammas}.    The difference
can be ascribed to the modified standardization.

\subsection{Fixed point}

Fixed point functions for the PDE above can be found by looking for
$s$-independent solutions $h(q,s)\equiv f(q)$ with a constant
$\gamma(s)\equiv \gamma$. From \eqref{eq:pde-clt} we obtain
\be
q\, \gamma\, f'(q) = 1,
\ee
whence the solution for $q<0$, satisfying the standardization
$f(-1)=0$, is
\be
f(q;\gamma) = \frac{1}{\gamma} \ln (-q).\label{eq:fp-clt}
\ee
Hence the moment generating function in the fixed point is by
\eqref{eq:def-h-clt}
\be
\Phi(q;\gamma) = e^{-(-q)^{-\frac{1}{\gamma}}}.
\ee
This is precisely of the form of the FTW distribution with $\gamma<0$ originally
obtained by Fisher and Tippett \cite{FisherTippett:1928}.  Now this function
corresponds to the moment generating function of the stable distributions with
even symmetry, usually written as
\be
\Phi_\alpha(q) = e^{-|q|^{\alpha}},
\ee
with $0<\alpha\leq 2$. By identification, we thus get $\alpha=-1/\gamma$.
The restriction for $\alpha$ is imposed to have a valid
probability density in the relation \eqref{eq:mom-gen}, $\alpha=2$ corresponding
to a Gaussian and $\alpha<2$ to the L\'evy family.  Note that the validity of
the moment generating function (namely, its inverse Fourier transform should be
a positive function) is in itself not ensured for any solutions of the
PDE \eqref{eq:pde-clt}, but the PDE preserves validity if the initial condition
was appropriate.

In conclusion, we obtained the even limit distributions for the sum of i.i.d.\
random variables in terms of their moment generating function, which is on the
negative semi-axis identical to the FTW family of limit distributions in EVS.
The correspondence holds with $\alpha=-1/\gamma$, where $\alpha$ is the L\'evy
index and $\gamma$ the parameter of the FTW distribution, in the region
$\gamma\leq -1/2$.

\subsection{Eigenfunctions}

Again in analogy to the EVS problem, we can ask about the eigenfunctions of the
RG transformation near a fixed point.  Consider perturbing the fixed point
function \eqref{eq:fp-clt}, whose second argument $\gamma$ we shall omit while
assuming it fixed, as
\be
h(q,s) = f\big( q+\epsilon(s)\,\psi(q)\big) \approx
f(q)+\epsilon(s) f'(q) \,\psi(q).\label{eq:pert-clt}
\ee
The symmetry of $h(q,s)$ in $q$ requires the function $\psi(q)$ to be odd, and  standardization conditions will be imposed as
\begin{subequations} \label{eq:pert-stand-clt}
 \begin{align}
  \psi(-1)&=0,\\
  \psi'(-1)&=-\frac{1}{\gamma},
 \end{align}
\end{subequations}
whereby from \eqref{eq:gamma-clt} we obtain
\be
\gamma(s) = \gamma + \epsilon(s)
\ee
Then substitution of \eqref{eq:pert-clt} into the PDE
\eqref{eq:pde-clt} and linear expansion in $\epsilon$ results in
\begin{equation}
 \dot \epsilon(s) \psi(q) = \epsilon(s)\big( q - \gamma\,\psi(q) +
q\,\gamma
\,\psi'(q)\big).
\end{equation}
This can be solved for nonzero $\epsilon$ only if
\begin{equation}
\frac{\dot \epsilon(s)}{\epsilon(s)}  = \gamma'
\end{equation}
is a constant, whence we obtain
\begin{equation}
\gamma' \psi(q) =  q - \gamma\,\psi(q) + q\,\gamma\, \psi'(q).
\end{equation}
The solution meeting the  conditions \eqref{eq:pert-stand-clt} is
straightforwardly obtained, for $q<0$ it is
\begin{equation}
 \psi(q;\gamma,\gamma')= \frac{1}{\gamma'}\left( q +
(-q)^{(\gamma+\gamma')/\gamma} \right).\label{eq:psi-clt}
\end{equation}
Extending for all $q$'s with odd symmetry, and using the L\'evy index $\alpha$ and the notation  $\delta=-\alpha\gamma'$ we eventually arrive at
\begin{equation}
 \psi_{\alpha,\delta}(q)= -\frac{\alpha q}{\delta}  \left( 1 -
|q|^{\delta} \right).
\end{equation}
Note the similarity of the eigenfunction with the one
obtained for the EVS problem \eqref{eq:psi-rg}: both are
essentially power functions with a linear function added to meet
standardization.  In particular, the function \eqref{eq:psi-clt} is
obtained from \eqref{eq:psi-rg} in the FTW case with a suitable shift
and rescaling.

The present treatise of the RG for the central limit problem is only
meant to be a demonstration of the close similarity with EVS. While it
can have further consequences for convergence to the limit, possible
exact solutions of the RG equation, and generalization to non-symmetric
distributions, we relegate those to further studies.

%%%%%%%%%%%%%%%%%%%%%%%%%%%%%%%%%%%%%%%%%%%%%%%%%%%%%%%%%%%%%%%%%%%%%%
%%%%%%
\section{Conclusion and outlook}
\label{S:conc}
%%%%%%%%%%%%%%%%%%%%%%%%%%%%%%%%%%%%%%%%%%%%%%%%%%%%%%%%%%%%%%%%%%%%%%
%%%%%%

The RG transformation for the EVS problem can be considered a simple if
not the simplest nonlinear RG conceivable:  it consists of raising the
integrated distribution function to a power and applying a linear change
of the random variable.  This gave rise to a linear, first order PDE,
albeit with a parametric driving determined by the parent distribution,
or equivalently, the initial condition. On the other hand, the
coefficient can also be viewed as the curvature of the solution in the
origin, which makes the PDE nonlinear but without an external forcing
term. The versatility of the PDE is reflected by the fact that it can
also account for discontinuities in the integrated distribution
function.

The PDE approach  allowed for a quite short and
straightforward derivation of the fixed line of limit distributions and
of the eigenfunctions. Furthermore, we found various exact solutions for
the RG path in terms of fixed points and eigenfunctions, i.e., invariant
manifolds, valid arbitrarily far from the fixed point. In general, we
found that all solutions, extending also to the nonlinear region away
from the fixed line, can be interpreted as linear combinations  of
eigenfunctions with possibly continuous weights.  We suggest that this
peculiar structure of function space, unparalleled in other RG methods
in statistical physics, is the consequence of the extreme simplicity of
the RG operation.  An analogous RG operation in terms of a PDE could
also be defined for the central limit problem, but on the moment
generating function rather than the integrated distribution, and it
allowed the simple rederivation of the known, even, L\'evy stable
functions as well as the calculation of eigenfunctions.

There is a plethora of further directions worth pursuing. The most
obvious application of the RG theory, not discussed in this paper, is to
calculate finite size effects (see \cite{GyorgyiETAL:2010}).  While the
limit distributions for the EVS problem of two random variables are
known, having more intricate properties than the above studied single
variable case, an RG approach there may be useful to reveal finite size
effects. Other natural extensions to the statistics of higher order
maxima and order statistics in general are also conceivable.   Whereas
there have been many advances thus far for the EVS of various problems
of correlated variables, especially in random walks, extensions of  the
RG concept again may prove to be useful. In fact, in a multitude of
problems in the broader field of front propagation, like random
fragmentation, extremal paths on trees, and randomly generated binary
trees, functional recursions and fixed point conditions are known.
Extensions of them to appropriate RG pictures and their eigenvalue
analysis holds the potential that finite size relaxation effects can be
described in terms of eigenfunctions. Another generalization of the EVS
paradigm for i.i.d.\ variables exists to cases where the batch size is
large but is itself a random variable.  Here the finite size effects can
presumably be studied again with the help of the RG results. Finally,
the formal analogy with the central limit problem holds the promise of a
simple approach to finite size effects there as well.

\begin{acknowledgments}
Kind hospitality by M.\ Droz on multiple occasions and
illuminating discussions with him and Z.\ R\'acz are hereby
gratefully acknowledged. E.~B. is pleased to thank the
E\"otv\"os University where part of this work was done during
his visit. This research has been partly supported by the Swiss NSF
and by Hungarian OTKA Grants No.\ T043734, K75324.
\end{acknowledgments}

\appendix
\section{Relation between alternative forms of the RG equations}
\label{S:ustable-ftg2-4}

Whereas this paper presents, and focuses on, the differential representation of
the RG transformation for EVS, it is insightful to perform a direct,
non-differential treatment.  Thus we reanalyze the relationship of
the parameterization \eqref{eq:sol-gen-nonlin} with the original, defining
formula of the RG transformation, which is essentially based on shift and
rescaling operations.  This way we again demonstrate that the two formulations
are equivalent.  We begin our reasoning by reconsidering the
linearized RG transform by a direct formulation.

So far in the paper, linear analysis of the RG has been based
on the linearization of the PDE. An alternative approach
is to linearize directly the RG transform \eqref{eq:gxs}, which was followed in
\cite{GyorgyiETAL:2008,GyorgyiETAL:2010} in the calculation of eigenfunctions,
a treatment we develop below.  We consider a perturbation $\phi(x,s)$, like in
Sec.\ \ref{S:ustable-ftg2-1},  defined by
\be
g(x,s)=x+\phi(x,s), \label{eq:g-phi-2}
\ee
the corresponding parent being denoted as
\be
g(x)=x+\phi_0(x) \label{eq:g-par}
\ee
Starting from \eqref{eq:gxs} we get for the RG transform
\be
x + \phi(x,s) = a(s)x+b(s)+\phi_0\big(a(s)x+b(s)\big)-s
\label{cn-RG0}
\ee
Assuming that $\phi=\phi_\mathrm{lin}$ is small, one gets by linearizing
\eqref{eq:bs} that $b(s)=s-\phi_0(s)$. Similarly, \eqref{eq:a-bdot} yields
$a(s)=\dot{b}(s) =1-\phi_0'(s)$.
Inserting these expressions for $a(s)$ and $b(s)$ in the RG \eqref{cn-RG0},
and linearizing the resulting equation yields
\be
\phi_\mathrm{lin}(x,s) = \phi_0(x+s)-x\phi_0'(s)-\phi_0(s).
\label{cn-RGlin0}
\ee
This is precisely the solution \eqref{eq:phi-ftg} of the linearized PDE
in the FTG case, with $\phi_0(x)=-\chi(x)$.
Hence this solution can straightforwardly be obtained by linearizing the
RG \eqref{eq:gxs}, rather than by linearizing the PDE and then looking for the
generic solution.

The linearized RG transform \eqref{cn-RGlin0} can be reformulated
as
\be
\phi_\mathrm{lin}(x,s) = \phi_0(x+s) + \bar{a}(s)x+\bar{b}(s),
\label{cn-RGlin1}
\ee
where $\bar{a}(s)$ and $\bar{b}(s)$ determine the shift by a linear function, on
$\phi_\mathrm{lin}$ that allows it to fulfill the standardization conditions
\be
\phi_\mathrm{lin}(0,s)=0, \qquad \partial_x\phi_\mathrm{lin}(0,s) = 0,
\label{cn-eqstd}
\ee
yielding
% \begin{subequations}
% \begin{align}
%  \bar{a}(s)&=-\phi_0'(s)   \label{eq:cn-bara}\\
% \bar{b}(s)&=-\phi_0(s).\label{eq:cn-barb}
% \end{align}
% \end{subequations}
\be \bar{a}(s)=-\phi_0'(s), \qquad
\bar{b}(s)=-\phi_0(s).\label{eq:cn-barb}\ee
In this way, the linearized RG clearly appears as a (constant-speed)
translation of the variable $x$, followed by a shift by a linear function
in order to restore the standardization conditions.

Let us now see how we can get the solution of the full RG from the linearized
one in this framework.  To this aim, we first reformulate the full RG in terms
of the function $\phi(x,s)$ defined in \eqref{eq:g-phi-2}, but this time without
making any approximation. Rearranging  \eqref{cn-RG0} gives
\be
\phi(x,s) = \phi_0\big(a(s)x+b(s)\big) + \big(a(s)-1\big)x + \big(b(s)-s\big),
\label{cn-RG1}
\ee
as the general form of an RG trajectory.

Now let us use $\phi_\mathrm{lin}(x,s)$  obeying \eqref{cn-RGlin1} and define a
new trial function $\phi(x,s)$ through a rescaling of $x$ and
a reparametrization of $s$, in the following way:
\be
\phi(x,s) \equiv \phi_\mathrm{lin}\big(\beta(s)x,\eta(s)\big),
\label{eq:phi-philin}
\ee
where the auxiliary functions $\beta(s),\eta(s)$ will be determined later.
Note that this trial function satisfies the standardization conditions
\eqref{cn-eqstd}. We now wish to check whether $\phi(x,s)$ could possibly
be a solution of the full RG \eqref{cn-RG1}.  One has
\be
\phi(x,s) = \phi_0(\beta x+\eta) + \bar{a}(\eta)\beta x+\bar{b}(\eta).
\ee
Comparison with \eqref{cn-RG1} leads us to set the  auxiliary functions
as $\beta(s)=a(s)$ and $\eta(s)=b(s)$.
% \begin{subequations}\label{eq:cn-ab}
% \begin{align}
% \eta(s)&=b(s),\label{eq:cn-etab} \\
% \beta(s)&=a(s).  \label{eq:cn-betaa}
% \end{align}
% \end{subequations}
Then the relations
\begin{subequations}\label{eq:cn-eqab}
\begin{align}
a(s)-1 &= \bar{a}\big(\eta(s)\big)\beta(s), \label{eq:cn-eqa} \\
b(s) - s &= \bar{b}\big(\eta(s)\big), \label{eq:cn-eqb}
\end{align}
\end{subequations}
should also be satisfied, which is not a priori obvious.
However, these relations define the  shift of $\phi$ by a linear function, which
ensures that $\phi$ fulfills the standardization condition. As the trial
function $\phi$ by construction  meets this standardization, \eqref{eq:cn-eqab}
should necessarily be satisfied.  This can also be checked explicitly as
follows.  On the one hand, since $\eta=b$, we know that $g(\eta)=s$, or from
\eqref{eq:g-par} equivalently
\be
\eta+\phi_0(\eta)=s. \label{eq:eta-phi-s}
\ee
But this equation, when we take into account \eqref{eq:cn-barb}, turns out to
be just \eqref{eq:cn-eqb}. Next, taking the derivative of \eqref{eq:eta-phi-s}
with respect to $s$ yields $\dot\eta+\phi_0'(\eta)\dot\eta=1$.
Recalling (\ref{eq:a-bdot}) we have $a=\dot{b}$, that is,
$\beta=\dot\eta$, leading to \be \beta+\phi_0'(\eta)\beta=1, \ee which, by
our using \eqref{eq:cn-barb}, turns out to be identical to \eqref{eq:cn-eqa}.

So we can conclude that the rather
intuitively constructed trial function  \eqref{eq:phi-philin} indeed
incorporates the most general formula  \eqref{cn-RG1} for the RG path, just what
we wanted to show.

%\bibliography{refs_evs_v6}

\begin{thebibliography}{26}
\expandafter\ifx\csname natexlab\endcsname\relax\def\natexlab#1{#1}\fi
\expandafter\ifx\csname bibnamefont\endcsname\relax
  \def\bibnamefont#1{#1}\fi
\expandafter\ifx\csname bibfnamefont\endcsname\relax
  \def\bibfnamefont#1{#1}\fi
\expandafter\ifx\csname citenamefont\endcsname\relax
  \def\citenamefont#1{#1}\fi
\expandafter\ifx\csname url\endcsname\relax
  \def\url#1{\texttt{#1}}\fi
\expandafter\ifx\csname urlprefix\endcsname\relax\def\urlprefix{URL }\fi
\providecommand{\bibinfo}[2]{#2}
\providecommand{\eprint}[2][]{\url{#2}}

\bibitem[{\citenamefont{Bouchaud and M\'ezard}(1997)}]{BouchaudMezard:1997}
\bibinfo{author}{\bibfnamefont{J.-P.} \bibnamefont{Bouchaud}} \bibnamefont{and}
  \bibinfo{author}{\bibfnamefont{M.}~\bibnamefont{M\'ezard}},
  \bibinfo{journal}{J. Phys. A} \textbf{\bibinfo{volume}{30}},
  \bibinfo{pages}{7997} (\bibinfo{year}{1997}).

\bibitem[{\citenamefont{Gy\"{o}rgyi et~al.}(2003)\citenamefont{Gy\"{o}rgyi,
  Holdsworth, Portelli, and R\'{a}cz}}]{GyorgyiETAL:2003}
\bibinfo{author}{\bibfnamefont{G.}~\bibnamefont{Gy\"{o}rgyi}},
  \bibinfo{author}{\bibfnamefont{P.~C.~W.} \bibnamefont{Holdsworth}},
  \bibinfo{author}{\bibfnamefont{B.}~\bibnamefont{Portelli}}, \bibnamefont{and}
  \bibinfo{author}{\bibfnamefont{Z.}~\bibnamefont{R\'{a}cz}},
  \bibinfo{journal}{Phys. Rev. E} \textbf{\bibinfo{volume}{68}},
  \bibinfo{pages}{056116} (\bibinfo{year}{2003}).

\bibitem[{\citenamefont{{Le Doussal} and
  Monthus}(2003)}]{LeDoussalMonthus:2003}
\bibinfo{author}{\bibfnamefont{P.}~\bibnamefont{{Le Doussal}}}
  \bibnamefont{and} \bibinfo{author}{\bibfnamefont{C.}~\bibnamefont{Monthus}},
  \bibinfo{journal}{Physica A} \textbf{\bibinfo{volume}{317}},
  \bibinfo{pages}{140} (\bibinfo{year}{2003}).

\bibitem[{\citenamefont{Majumdar and Comtet}(2004)}]{MajumdarComtet:2004}
\bibinfo{author}{\bibfnamefont{S.~N.} \bibnamefont{Majumdar}} \bibnamefont{and}
  \bibinfo{author}{\bibfnamefont{A.}~\bibnamefont{Comtet}},
  \bibinfo{journal}{Phys. Rev. Lett.} \textbf{\bibinfo{volume}{92}},
  \bibinfo{pages}{225501} (\bibinfo{year}{2004}).

\bibitem[{\citenamefont{Schehr and Majumdar}(2006)}]{SchehrMajumdar:2006}
\bibinfo{author}{\bibfnamefont{G.}~\bibnamefont{Schehr}} \bibnamefont{and}
  \bibinfo{author}{\bibfnamefont{S.~N.}~\bibnamefont{Majumdar}},
  \bibinfo{journal}{Phys. Rev. E} \textbf{\bibinfo{volume}{73}},
  \bibinfo{pages}{056103} (\bibinfo{year}{2006}).

\bibitem[{\citenamefont{Krapivsky and Majumdar}(2000)}]{KrapivskyMajumdar:2000}
\bibinfo{author}{\bibfnamefont{P.~L.} \bibnamefont{Krapivsky}}
  \bibnamefont{and} \bibinfo{author}{\bibfnamefont{S.~N.}
  \bibnamefont{Majumdar}}, \bibinfo{journal}{Phys. Rev. Lett.}
  \textbf{\bibinfo{volume}{85}}, \bibinfo{pages}{5492} (\bibinfo{year}{2000}).

\bibitem[{\citenamefont{Gumbel}(1958)}]{Gumbel:1958}
\bibinfo{author}{\bibfnamefont{E.~J.} \bibnamefont{Gumbel}},
  \emph{\bibinfo{title}{Statistics of Extremes}} (\bibinfo{publisher}{Dover
  Publications}, \bibinfo{year}{1958}).

\bibitem[{\citenamefont{Katz et~al.}(2002)\citenamefont{Katz, Parlange, and
  Naveau}}]{KatzParlangeNaveau:2002}
\bibinfo{author}{\bibfnamefont{R.~W.} \bibnamefont{Katz}},
  \bibinfo{author}{\bibfnamefont{M.~B.} \bibnamefont{Parlange}},
  \bibnamefont{and} \bibinfo{author}{\bibfnamefont{P.}~\bibnamefont{Naveau}},
  \bibinfo{journal}{Adv.~Water~Resour.} \textbf{\bibinfo{volume}{25}},
  \bibinfo{pages}{1287} (\bibinfo{year}{2002}).

\bibitem[{\citenamefont{Sornette et~al.}(1996)\citenamefont{Sornette, Knopoff,
  Kagan, and Vannest}}]{SornetteETAL:1996}
\bibinfo{author}{\bibfnamefont{D.}~\bibnamefont{Sornette}},
  \bibinfo{author}{\bibfnamefont{L.}~\bibnamefont{Knopoff}},
  \bibinfo{author}{\bibfnamefont{Y.}~\bibnamefont{Kagan}}, \bibnamefont{and}
  \bibinfo{author}{\bibfnamefont{C.}~\bibnamefont{Vannest}},
  \bibinfo{journal}{J. Geophys. Res.} \textbf{\bibinfo{volume}{101}},
  \bibinfo{pages}{13883} (\bibinfo{year}{1996}).

\bibitem[{\citenamefont{Embrecht et~al.}(1997)\citenamefont{Embrecht,
  Kl\"uppelberg, and Mikosch}}]{EmbrechtETAL:1997}
\bibinfo{author}{\bibfnamefont{P.}~\bibnamefont{Embrecht}},
  \bibinfo{author}{\bibfnamefont{C.}~\bibnamefont{Kl\"uppelberg}},
  \bibnamefont{and} \bibinfo{author}{\bibfnamefont{T.}~\bibnamefont{Mikosch}},
  \emph{\bibinfo{title}{Modelling {E}xtremal {E}vents for {I}nsurance and
  {F}inance}} (\bibinfo{publisher}{Springer, Berlin}, \bibinfo{year}{1997}).

\bibitem[{\citenamefont{Longin}(2000)}]{Longin:2000}
\bibinfo{author}{\bibfnamefont{F.}~\bibnamefont{Longin}}, \bibinfo{journal}{J.
  Bank. Finance} \textbf{\bibinfo{volume}{24}}, \bibinfo{pages}{1097}
  (\bibinfo{year}{2000}).

\bibitem[{\citenamefont{Fisher and Tippett}(1928)}]{FisherTippett:1928}
\bibinfo{author}{\bibfnamefont{R.~A.} \bibnamefont{Fisher}} \bibnamefont{and}
  \bibinfo{author}{\bibfnamefont{L.~H.~C.} \bibnamefont{Tippett}},
  \bibinfo{journal}{Procs.~Cambridge~Philos.~Soc.}
  \textbf{\bibinfo{volume}{24}}, \bibinfo{pages}{180} (\bibinfo{year}{1928}).

\bibitem[{\citenamefont{Gnedenko}(1943)}]{Gnedenko:1943}
\bibinfo{author}{\bibfnamefont{B.~V.} \bibnamefont{Gnedenko}},
  \bibinfo{journal}{Ann.~Math.} \textbf{\bibinfo{volume}{44}},
  \bibinfo{pages}{423} (\bibinfo{year}{1943}).

\bibitem[{\citenamefont{Galambos}(1978)}]{Galambos:1978}
\bibinfo{author}{\bibfnamefont{J.}~\bibnamefont{Galambos}},
  \emph{\bibinfo{title}{The {A}symptotic {T}heory of {E}xtreme {V}alue
  {S}tatistics}} (\bibinfo{publisher}{John Wiley \& Sons},
  \bibinfo{year}{1978}).

\bibitem[{\citenamefont{von Mises}(1954)}]{vonMises:1954}
\bibinfo{author}{\bibfnamefont{R.}~\bibnamefont{von Mises}},
  \bibinfo{journal}{Amer. Math. Soc. (Providence, R.I.)}
  (\bibinfo{year}{1954}).

\bibitem[{\citenamefont{Clusel and Bertin}(2008)}]{CluselBertin:2008}
\bibinfo{author}{\bibfnamefont{M.}~\bibnamefont{Clusel}} \bibnamefont{and}
  \bibinfo{author}{\bibfnamefont{E.}~\bibnamefont{Bertin}},
  \bibinfo{journal}{Int.~J.~Mod.~Phys.~B} \textbf{\bibinfo{volume}{22}},
  \bibinfo{pages}{3311} (\bibinfo{year}{2008}).

\bibitem[{\citenamefont{Hall}(1979)}]{Hall:1979}
\bibinfo{author}{\bibfnamefont{P.}~\bibnamefont{Hall}},
  \bibinfo{journal}{Journal of Applied Probability}
  \textbf{\bibinfo{volume}{16}}, \bibinfo{pages}{433} (\bibinfo{year}{1979}).

\bibitem[{\citenamefont{Hall}(1980)}]{Hall:1980}
\bibinfo{author}{\bibfnamefont{P.}~\bibnamefont{Hall}},
  \bibinfo{journal}{Advances in Applied Probability}
  \textbf{\bibinfo{volume}{12}}, \bibinfo{pages}{491} (\bibinfo{year}{1980}).

\bibitem[{\citenamefont{de~Haan and Resnick}(1996)}]{DeHaanResnick:1996}
\bibinfo{author}{\bibfnamefont{L.}~\bibnamefont{de~Haan}} \bibnamefont{and}
  \bibinfo{author}{\bibfnamefont{S.}~\bibnamefont{Resnick}},
  \bibinfo{journal}{Annals of Probability} \textbf{\bibinfo{volume}{24}},
  \bibinfo{pages}{97} (\bibinfo{year}{1996}).

\bibitem[{\citenamefont{de~Haan and Stadtm{\"u}ller}(1996)}]{DeHaanStadm:1996}
\bibinfo{author}{\bibfnamefont{L.}~\bibnamefont{de~Haan}} \bibnamefont{and}
  \bibinfo{author}{\bibfnamefont{U.}~\bibnamefont{Stadtm{\"u}ller}},
  \bibinfo{journal}{J. Australian Math. Soc.} \textbf{\bibinfo{volume}{61}},
  \bibinfo{pages}{381} (\bibinfo{year}{1996}).

\bibitem[{\citenamefont{Gy\"{o}rgyi et~al.}(2008)\citenamefont{Gy\"{o}rgyi,
  Moloney, Ozog\'{a}ny, and R\'{a}cz}}]{GyorgyiETAL:2008}
\bibinfo{author}{\bibfnamefont{G.}~\bibnamefont{Gy\"{o}rgyi}},
  \bibinfo{author}{\bibfnamefont{N.~R.} \bibnamefont{Moloney}},
  \bibinfo{author}{\bibfnamefont{K.}~\bibnamefont{Ozog\'{a}ny}},
  \bibnamefont{and} \bibinfo{author}{\bibfnamefont{Z.}~\bibnamefont{R\'{a}cz}},
  \bibinfo{journal}{Phys. Rev. Lett.} \textbf{\bibinfo{volume}{100}},
  \bibinfo{pages}{210601} (\bibinfo{year}{2008}).

\bibitem[{\citenamefont{Gy\"{o}rgyi et~al.}(2010)\citenamefont{Gy\"{o}rgyi,
  Moloney, Ozog\'{a}ny, R\'{a}cz}, and Droz}]{GyorgyiETAL:2010}
\bibinfo{author}{\bibfnamefont{G.}~\bibnamefont{Gy\"{o}rgyi}},
  \bibinfo{author}{\bibfnamefont{N.~R.} \bibnamefont{Moloney}},
  \bibinfo{author}{\bibfnamefont{K.}~\bibnamefont{Ozog\'{a}ny}},
  \bibinfo{author}{\bibfnamefont{Z.}~\bibnamefont{R\'{a}cz}},
  \bibnamefont{and} \bibinfo{author}{\bibfnamefont{M.}~\bibnamefont{Droz}},
  \bibinfo{journal}{Phys. Rev. E} \textbf{\bibinfo{volume}{81}},
  \bibinfo{pages}{041135} (\bibinfo{year}{2010}).

\bibitem[{\citenamefont{Zinn-Justin}(2007)}]{ZinnJustin:2005}
\bibinfo{author}{\bibfnamefont{J.}~\bibnamefont{Zinn-Justin}},
  \emph{\bibinfo{title}{Phase Transitions and Renormalisation Group}}
  (\bibinfo{publisher}{Oxford University Press}, \bibinfo{year}{2007}).

\bibitem[{\citenamefont{Jona-Lasinio}(2001)}]{JonaLasinio:2001}
\bibinfo{author}{\bibfnamefont{G.}~\bibnamefont{Jona-Lasinio}},
  \bibinfo{journal}{Phys. Rep.} \textbf{\bibinfo{volume}{352}},
  \bibinfo{pages}{439} (\bibinfo{year}{2001}).

\bibitem[{\citenamefont{Schehr and {Le Doussal}}(2010)}]{SchehrLeDoussal:2009}
\bibinfo{author}{\bibfnamefont{G.}~\bibnamefont{Schehr}} \bibnamefont{and}
  \bibinfo{author}{\bibfnamefont{P.}~\bibnamefont{{Le Doussal}}},
  \bibinfo{journal}{Journal of Statistical Mechanics: Theory and Experiment}
  \textbf{\bibinfo{volume}{2010}}, \bibinfo{pages}{P01009}
  (\bibinfo{year}{2010}).

\bibitem[{\citenamefont{Gnedenko and Kolmogorov}(1954)}]{Kolmogorov:1954}
\bibinfo{author}{\bibfnamefont{B.~V.} \bibnamefont{Gnedenko}} \bibnamefont{and}
  \bibinfo{author}{\bibfnamefont{A.~N.} \bibnamefont{Kolmogorov}},
  \emph{\bibinfo{title}{Limit {D}istributions for {S}ums of {I}ndependent
  {R}andom {V}ariables}}, Addison-Wesley series in Mathematics
  (\bibinfo{publisher}{Addison-Wesley, Cambridge, MA}, \bibinfo{year}{1954}).

\bibitem[{\citenamefont{Christoph and Wolf}(1993)}]{Christoph:1993}
\bibinfo{author}{\bibfnamefont{G.}~\bibnamefont{Christoph}} \bibnamefont{and}
  \bibinfo{author}{\bibfnamefont{W.}~\bibnamefont{Wolf}},
  \emph{\bibinfo{title}{Convergence {T}heorems with a {S}table {L}imit {L}aw}}
  (\bibinfo{publisher}{Akademie-Verlag, Berlin}, \bibinfo{year}{1993}).



\end{thebibliography}
\end{document}